%% file: Litvinov_et_al_CatchSlip.tex
\documentclass[fleqn,10pt]{wlscirep}
\usepackage{multibib}
\usepackage{appendix}
\usepackage{multirow}
\usepackage{xcolor}
\usepackage{colortbl}

\title{Molecular Mechanism of Transition from Catch-Bond to Slip-Bond in Fibrin}

\author[1,2]{Rustem I. Litvinov}
\author[3,4, $\P$]{Olga Kononova}
\author[3]{Farkhad Maksudov}
\author[4]{Artem Zhmurov}
\author[3]{Kenneth A. Marx}
\author[1,*]{John W. Weisel}
\author[3,4,*]{Valeri Barsegov}
\affil[1]{Department of Cell and Developmental Biology, University of Pennsylvania School of Medicine, Philadelphia, PA 19104, USA}
\affil[2]{Institute of Fundamental Medicine and Biology, Kazan Federal University, Kazan 420008, Russian Federation}
\affil[3]{Department of Chemistry, University of Massachusetts, Lowell, MA 01854, USA}
\affil[4]{Moscow Institute of Physics and Technology, Moscow Region 141700, Russian Federation}

\affil[*]{weisel@mail.med.upenn.edu, valeri\_barsegov@uml.edu} 

\affil[$\P$]{Current address: Department of Material Science and Engineering, University of California, Berkeley, Berkeley, CA, 94720, USA} 

\keywords{catch-slip bond, fibrin polymerization, fluctuating bottleneck, molecular modeling; Molecular Dynamics simulations on a GPU}

\begin{abstract}
The lifetimes of non-covalent A:a knob-hole bonds in fibrin probed with the optical trap-based force-clamp first increases (``catch bonds'') and then decreases (``slip bonds'') with increasing tensile force. 
Molecular modeling of ``catch-to-slip'' transition using the atomic structure of the A:a complex reveals that the movable flap serves as tension-dependent molecular switch. 
Flap dissociation from the regulatory B-domain in $\gamma$-nodule and translocation from the periphery to knob `A' triggers the hole `a' closure and interface remodeling, which results in the increased binding affinity and prolonged bond lifetimes. 
Fluctuating bottleneck theory is developed to understand the ``catch-to-slip'' transition in terms of the interface stiffness $\kappa =$ 15.7 pN nm $^{-1}$, interface size fluctuations 0.7-2.7 nm, knob `A' escape rate constant $k_0 =$ 0.11 nm$^2$ s$^{-1}$, and transition distance for dissociation $\sigma_y =$ 0.25 nm. 
Strengthening of the A:a knob-hole bonds under small tension might favor formation and reinforcement of nascent fibrin clots under hydrodynamic shear.
\end{abstract}

\begin{document}

\flushbottom
\maketitle
\thispagestyle{empty}

%\noindent Please note: Abbreviations should be introduced at the first mention in the main text – no abbreviations lists. Suggested structure of main text (not enforced) is provided below.

\section*{Introduction}

Fibrin is the end product of blood clotting that constitutes a proteinaceous 3D network providing a filamentous mechanical scaffold of clots and thrombi. 
Formation of fibrin is essential for hemostasis, thrombosis, and antimicrobal host defense; fibrin is also widely used as a biomaterial. \cite{Litvinov&WeiselSTH16} 
Fibrin is formed from a blood plasma protein, fibrinogen, which converts to monomeric fibrin that polymerizes to form soluble fibrin protofibrils. 
These further elongate and aggregate laterally into insoluble thick fibers that branch to form the space-filling network (Fig.~\ref{fig:fig1}a-c). 
During and after polymerization fibrin is covalently cross-linked by a plasma transglutaminase, factor XIIIa, that makes fibrin stiff and resistant to enzymatic lysis (Fig.~\ref{fig:fig1}a-c).\cite{Weisel&LitvinovSB17} 
Mechanical stability of nascent blood clots in response to forces imposed by the blood flow, contracting platelets, and other dynamic factors is determined by the strength of the knob-hole interactions prior to cross-linking by factor XIIIa. 
Consequently, the dynamics of association-dissociation transitions in knob-hole complexes govern formation of fibrin, influence the final structure and mechanical stability of clots and thrombi, including clot rupture (embolization) and shrinkage (contraction, retraction). 
Impaired knob-hole interactions result in loose, weak, unstable clots and are associated with the tendency to bleed. 
Dense fibrin networks show increased stiffness, higher fibrinolytic resistance and mechanical resilience, which may predispose individuals to thrombotic cardiovascular diseases, such as heart attack and ischemic stroke.\cite{StandevenBR05}

Fibrinogen, the soluble fibrin precursor, is a 340-kDa protein comprising three pairs of polypeptide chains, A$\alpha$, B$\beta$, and $\gamma$, arranged into one central globular E region and two lateral globular D regions connected with $\alpha$-helical coiled coils (Fig.~\ref{fig:fig1}a). 
Thrombin splits off two pairs of fibrinopeptides A and B from the N-termini of the fibrinogen's A$\alpha$ and B$\beta$ chains, respectively, in the central E region. 
This results in the exposure of binding sites (knobs) `A' and `B' that interact with constitutively accessible complementary sites (holes) `a' and `b' located in the $\gamma$- and $\beta$-nodules, respectively, of the lateral D regions of another fibrin molecule (Fig.~\ref{fig:fig1}a-c).\cite{LaudanoPNAS78, LaudanoBiochemistry80, Weisel&LitvinovSB17} 
X-ray crystallographic studies of fibrinogen fragments revealed that holes `a' and `b' interact with the peptides GPRP and GHRP that mimic knobs `A' and `B', respectively, which comprise the newly exposed N-terminal motifs of the $\alpha$ and $\beta$ chains of fibrin.\cite{LaudanoPNAS78, SpraggonNature97} 
Binding of knobs `A' to holes `a' is necessary for fibrin polymerization, while the B:b bonds play a secondary role.\cite{Weisel&LitvinovSB17}

We employed single-molecule forced unbinding assays to probe the strength of A:a knob-hole bond \cite{LitvinovBlood05, LitvinovJTH13} which revealed unusual strengthening with the increasing pulling force applied to disrupt the bond, but the nature of this finding remained unclear. 
Such counterintuitive behavior has been described in the literature for a number of receptor-ligand pairs as the ``catch'' bond \cite{HertigCurrBiol12, MarshallNat03} in contrast to the commonly known ``slip'' bond that dissociates faster with the increasing force. 
Interestingly, a non-covalent bond can behave as a catch bond at low forces (typically $<$30-40 pN) and as a slip bond at higher forces, thus displaying the dual ``catch-slip'' character. 
Several receptor-ligand complexes showing the catch-slip transition have been characterized including coupled cell adhesion molecules and glycoprotein ligand-1,\cite{MarshallNat03} E-cadherin dimer,\cite{RakshitaPNAS12} integrin $\alpha$5$\beta$1 and fibronectin,\cite{KongJCB09} bacterial adhesin FimH,\cite{ThomasCell02, SauerNC16} von Willebrand factor and receptor GP1b,\cite{YagoJCI08, FeghhiBJ16} actomyosin,\cite{GuoPNAS06} microtubule-kinetochore,\cite{AkiyoshiNature10} and microtubule-dynein \cite{RaiCell13,NairPRE} complexes. 
The catch-slip phenomenon was studied experimentally \cite{YagoJCI08, ChenJBC10, McEverARCDB10, SauerNC16, WaldronPNAS08, SarangapaniJBC11} and computationally.\cite{HelmsFEBSLet16, SauerNC16, ManibogNC14, GunnersonJPCB09, ChakrabartiJSB17, Vernerey&AkalpPRE16, Bullerjahn&KroyPRE16} 
Theoretical models have been purposed including the two-state model, \cite{BarsegovPNAS05, BarsegovJPCB06} two-pathway model,\cite{PereverzevBJ05} sliding-rebinding model,\cite{LouBJ07} hydrogen bond network model,\cite{ChakrabartiPNAS14} and other models.\cite{LiuPRE06, WeiPRE08} 
Yet, the atomic-level structural basis underlying the catch-slip transition in receptor-ligand complexes has eluded detailed characterization \cite{HertigCurrBiol12}.

Here, we combined the single-molecule forced unbinding experiments \textit{in vitro} and \textit{in silico} using nanomechanical measurements and Molecular Dynamics (MD) simulations to resolve the structural mechanism underlying the dual catch-slip response of fibrin polymers to tension. 
We show that the strength of A:a knob-hole bonds first increases with tensile force up to $f$ $\approx$ 30-35 pN (catch bond) and then decreases with force at $f$ $>$ 35 pN (slip bonds). 
Forced dissociation assays \textit{in silico} revealed dynamic remodeling of the A:a association interface, which results in a manifold of bound states with tension-dependent binding affinity. 
We developed new fluctuating bottleneck theory to model the experimental distributions of bond lifetimes and average bond lifetime as a function of force. 
The results provide a comprehensive structure-based interpretation of the experimentally observed catch-slip dynamics of dissociation of the A:a knob-hole bonds, the strongest non-covalent interactions in early stages of fibrin polymerization (Fig.~\ref{fig:fig1}). 
Our theory can be extended and generalized to characterize biomolecular interactions in receptor-ligand pairs that form deep binding pockets.

\section*{Results}

\subsection*{Dissociation kinetics of the A:a knob-hole bonds under constant tensile force}

\textbf{Bond lifetime of the fibrin-fibrinogen complex as a function of tensile force:} 
The A:a knob-hole interactions were reproduced at an interface during repeated touching of two microscopic beads coated covalently with fibrinogen or monomeric fibrin. 
The fibrin molecule is a source of knobs `a' and fibrinogen molecule is a source of holes `a'. 
By touching the fibrin-coated surface with the fibrinogen-coated surface, we allowed these molecules to associate forming the A:a knob-hole complex (binding phase). 
Next, the fibrin-coated and fibrinogen-coated surfaces were retracted by a constant tensile (pulling) force to dissociate the A:a knob-hole bond (unbinding phase). 
By repeating the binding-unbinding cycles, we were able to measure the times-to-dissociation (lifetimes) for the fibrin-fibrinogen complexes and to probe the dependence of A:a knob-hole bond lifetimes on a constant pulling force $f$. 
We varied the pulling force in the 5-60-pN-range and collected for each force value the distributions of bond lifetimes $P(\tau; f)$, which were then used to calculate the average bond lifetimes $\langle \tau \rangle$. 
We found that the fibrin-fibrinogen interactions display a non-monotonic dependence of $\langle \tau \rangle$ on $f$: $\langle \tau \rangle$  first increased with $f$ up to $f$ $=$ 30-35 pN and then decreased at $f$ $>$ 40 pN (Fig.~\ref{fig:fig2}a).

To distinguish the A:a knob-hole binding from other interactions, we carried out a number of control experiments. 
Weak non-specific surface-to-surface adhesion events produced interaction signals with the bond lifetimes $<$0.03s. 
Specificity was also tested by coating the interacting surfaces with an inert protein lacking knobs and holes (bovine serum albumin, BSA) or with a relevant protein lacking knobs `a' and `b' (fibrinogen). 
The control interactions (fibrinogen/BSA, fibrin/BSA, fibrinogen/fibrinogen interfaces) revealed mostly (93-97\%) short-lived attachment signals with bond lifetimes $<$0.03s and a small fraction (0.6-2.5\%) of more stable interactions lasting 0.04-0.5 s (Table~\ref{tab:tabS1}). 
The bond lifetimes of control interactions did not show any dependence on force (Fig.~\ref{fig:fig2}a). 
To distinguish the A:a knob-hole binding from other associations, we measured the fibrinogen-fibrin interaction in the absence and presence of the GPRPam peptide -- specific competitive knob `a' inhibitor. 
GPRPam suppressed interactions lasting $>$0.5 s (Fig.~\ref{fig:figS1}, Table~\ref{tab:tab1}), which were, therefore, considered to reflect the specific A:a knob-hole interactions. 
The binding probability for these interactions lasting $>$0.5s was several-fold higher in the fibrinogen-fibrin system than in control non-specific interactions described above, including fibrin-fibrinogen interactions inhibited by GPRP peptide (Fig.~\ref{fig:figS2}, Table~\ref{tab:tabS1}).

To test whether the intermediately strong interactions with bond lifetimes 0.03s $< \tau <$ 0.5s contribute to the observed dependence of $\langle \tau \rangle$ on $f$, we also compared the bond lifetimes obtained either by including or excluding these intermediate-strength interactions from the data sets. 
The profiles of $\langle \tau \rangle$ vs. $f$ (Fig.~\ref{fig:figS3}) as well as the cumulative binding probability vs. $f$ (Fig.~\ref{fig:figS4}) were quite similar with a corresponding small shift to shorter bond lifetimes (Fig.~\ref{fig:figS4}) and to a higher binding probability (Fig.~\ref{fig:figS3}). 
Based on these findings, we concluded that the bond lifetimes $\tau >$ 0.5s represent specific fibrin-fibrinogen or A:a knob-hole interactions. 
We included these data into subsequent statistical analyses and modeling, while the bond lifetimes $\tau <$ 0.5s were excluded from the data analysis.

\textbf{Bond lifetimes of the D:E complex as a function of tensile force:} 
To minimize the role of non-specific interactions due to large size of fibrin and fibrinogen molecules, we replaced the full-length fibrinogen with its smaller proteolytic fragment D containing mainly a lateral globular portion. 
Monomeric fibrin was replaced with fragment E comprising mainly the central globule. Fragment D bears one constitutively open hole `a'. 
Fragment E can exist in three variant, namely: i) it can bear both knobs `a' and `b' after cleaving of fibrinopeptides A and B with thrombin (called fragment desAB-E); ii) it can bear only knobs `a' after cleaving of fibrinopeptide A with batroxobin (called fragment desA-E) and iii) it can have no knobs if remains untreated (fragment E) (Fig.~\ref{fig:fig1}d). 
We found that specific interactions between fragments D and desAB-E also displayed dual catch-slip profile. The average bond lifetimes increased with force up to $f =$ 30-40 pN and then decreased at $f >$ 40 pN (Fig.~\ref{fig:fig2}b). 
The bell-like curve of $\langle \tau \rangle$ characteristic of the catch-to-slip transition was quite similar to the one obtained for the fibrinogen:fibrin interactions (Fig.~\ref{fig:fig2}b) albeit with shorter average bond-lifetimes. 
Importantly, the dual catch-slip behavior was largely suppressed with addition of GPRP (Fig.~\ref{fig:figS5}) and it disappeared altogether following a substantial reduction in surface density of fragment D (Fig.~\ref{fig:figS6}). 
These results strongly indicate that the D/desAB-E interactions are specific and that they reflect formation-dissociation of the knob-hole bonds. 
Exposure of knobs `a' in fragment desA-E (without knobs `b') preserves their ability to interact with fragment D in the catch-slip fashion. However, the average bond lifetimes were substantially shorter and the force-dependence of bond lifetimes was less pronounced (Fig.~\ref{fig:fig2}b). 
Intact fragment E with uncleaved fibrinopeptides A and B was non-reactive with fragment D (Fig.~\ref{fig:fig2}b). 
This provides yet another evidence that the measured knob-hole interactions of fragment D with fragments desAB-E and desA-E was specific and reflected the A:a knob-hole interactions.

\subsection*{Molecular Modeling of A:a knob-hole interactions}

\textbf{Dynamic force measurements \textit{in silico}:} 
We focused on the isolated A:a knob-hole complex using the structural model of A:a knob-hole complex as a part of reconstructed fibrin-fibrin D:E:D interactions based on the crystallographic data (Fig.~\ref{fig:fig1}e,g). 
Potentially, A:b interactions (knobs `a' binding to holes `b') and B:a interactions (knobs `b' binding to holes `a') might also affect the fibrinogen-fibrin complex lifetime, but our previous studies showed that the B:a interactions are unlikely to exist \cite{LitvinovBlood07} and that formation of the A:b knob-hole bonds is structurally and thermodynamically unfavorable.\cite{ZhmurovStructure16} 
Fibrinogen molecule has two pairs of strong Ca$^{2+}$-binding sites \cite{Weisel&LitvinovSB17} and the equilibrium dissociation constant for the high-affinity Ca$^{2+}$-binding sites in fibrinogen ($K_d$ $\sim$ 1 $\mu$M) implies that these sites are fully occupied with Ca$^{2+}$ in plasma environment. 
In our experiments, the catch-slip transition was observed both with and without Ca$^{2+}$ (data not shown), and so we did not include calcium ions in the computational modeling. 
We probed the strength of A:a knob-hole interactions using ramped force $f(t)$ (see Methods) with the pulling velocity $\nu_f = 10^3$ and $10^4$ $\mu$m/s. 
The distinct Pathway 1 (Pathway 2) of A:a knob-hole bond dissociation (Fig.~\ref{fig:figS7}a,b) is characterized by faster (slower) dissociation at lower (higher) molecular force $F$.
For the slower $\nu_f = 10^3$ $\mu$m/s, the bonds yielded at $F$ $\approx$ 60 pN in 50\% of 10 simulation runs (Pathway 1) and at $F$ $\approx$ 90 pN in 25\% of runs (Pathway 2), while in the remaining 25\% of trajectories the A:a knob-hole bonds dissociated at $F$ $\approx$ 70 pN (Fig.~\ref{fig:figS7}a). 
For the faster $\nu_f = 10^4$ $\mu$m/s, the A:a knob-hole bonds yielded at $F$ $\approx$ 90 pN (Pathway 1) and at $F$ $\approx$ 130 pN (Pathway 2) 70\% and 30\% of 10 trajectories, respectively (Fig.~\ref{fig:figS7}b).

\textbf{Binding affinity and maps of binding contacts:}
Next, we analyzed the simulations output. We monitored the dissociation dynamics by projecting the total number of persistent binding contacts $Q$ between the residues in hole `a' (in the $\gamma$-nodule ) and residues in knob `a' (in the $\alpha$ chain) as a function of time $t$. 
A pair of amino acids $i$ and $j$ is said to form a binary contact if the distance between the center-of-mass of their side chains $r_{ij} <$ 6.5 \AA~ persists for more than 10 ns. 
$Q$ is related to the binding affinity (i.e. more contacts means stronger binding) and it reflects instantaneous changes in the bond strength. 
Results are displayed in Fig.~\ref{fig:figS8}a (see also Fig.~\ref{fig:figS7}c, d). 
The profiles of $Q$ vs. $t$ show that at the beginning $Q$ $\approx$ 20 contacts (native bound state of the A:a complex), but with force ramping up $Q$ changes. 
For Pathways 1 resulting in the lower dissociation force $F$, $Q$ decays to zero with time. 
The moment of time $t = \tau$, at which $Q(\tau)=$ 0, marks complete dissociation of the A:a knob-hole bond. 
However, for Pathway 2 corresponding to higher values of $F$, the dynamics of $Q$ is non-monotonic (Fig.~\ref{fig:figS8}a and Fig.~\ref{fig:figS7}c, d): $Q$ initially increased to 30-35 contacts and then decreased to zero at longer times. 
Hence, the higher values of $F$ corresponding to Pathway 2 of the A:a knob-hole complex dissociation are directly correlated with the higher number of contacts $Q$, which is proportional to the knob-hole binding affinity.

To gather the residue-level information about the high- and low-affinity bound states of the A:a knob-hole complex, we analyzed entire maps of knob-hole contacts at the binding interface (residues $\gamma$Ser240--$\gamma$Lys380). 
The maps of binding contacts for the native bound state and for the intermediate states right before dissociation are compared in Fig.~\ref{fig:figS8}c,d for Pathway 1 and 2 trajectories. 
The bottom corner in the map corresponds to the native state, and the top corner corresponds to the intermediate state ($t =$ 0.1 $\mu$s). 
Dissociation along Pathway 1 is accompanied by a slight increase in contacts density between the movable flap (residues $\gamma$Phe295--Thr305) and $\beta$-sheet stack of the B-domain (residues $\gamma$Ile242--$\gamma$Gly283; compare the areas circled by solid and dashed ovals in Fig.~\ref{fig:figS8}c) and no substantial change in contacts density between the receptor and knob `a'. 
This correlates with decay in $Q$ for this trajectory (green curve in Fig.~\ref{fig:figS8}a). For Pathway 2, additional binding contacts form between the movable flap and knob `a', while the contacts between movable flap and $\beta$-sheet stack are disrupted (compare the areas circled by solid and dashed ovals in Fig.~\ref{fig:figS8}d). 
This is reflected by an increase in $Q$ for this trajectory (Fig.~\ref{fig:figS8}a). 
Hence, under the influence of pulling force, hole `a' transitions between two conformation types: one is characterized by weak interactions of movable flap with the $\beta$-sheet in B-domain leading to faster dissociation of the knob `a' at a lower unbinding force $F$ (Pathway 1); and the other type of conformation favors formation of additional binding contacts between the movable flap and knob `a', hence, prolonging the bond lifetime (Pathway 2).

\textbf{Low-affinity vs. high-affinity bound states:}
Analysis of binding affinity showed that in Pathway 2 the forced dissociation occurs from the high-affinity bound state compared to the low-affinity bound state observed in Pathway 1. 
We analyzed the simulation output to characterize conformational transitions that occur in the A:a knob-hole binding interface under tension. 
Fig.~\ref{fig:fig3} shows the time-dependent profiles of $F$ and $Q$ and corresponding atomic structures of the A:a knob-hole complex observed along dissociation Pathways 1-2. 
We found that the transition of the receptor-ligand complex from the low- to high-affinity bound states is controlled by the movable flap, which serves as a tension-sensitive ``molecular switch'', which triggers opening and closing of the binding interface. 
In Pathway 1 (Fig.~\ref{fig:fig3}), the moveable flap is far away from the ligand (snapshots 2a,3a in Fig.~\ref{fig:fig3}). 
Negatively charged residues $\gamma$Asp291, $\gamma$Asp294 and $\gamma$Asp297 interact with positively charged residues $\gamma$Arg256 and $\gamma$Arg275 of the $\beta$-sheet stack in B-domain. 
In some trajectories, we also observed a structural transition in the movable flap from the random coil to the $\beta$-strand, which then associates with the $\beta$-sheet stack of B-domain. 
As a result, the binding interface opens, facilitating dissociation of knob `a' from the low-affinity bound state. Corresponding to these observations an increase in the contact density between the movable flap and B-domain is manifest on the contact map (circled area in Fig.~\ref{fig:fig3}, left). 
In Pathway 2 (Fig.~\ref{fig:fig3}), the movable flap extends and translocates toward the ligand, forming additional binding contacts between residues $\gamma$Asp297, $\gamma$Asp298, $\gamma$Pro299, $\gamma$Ser300, $\gamma$Asp301, $\gamma$Lys302, $\gamma$Phe303, $\gamma$Phe304 in the flap and residues $\alpha$Gly17, $\alpha$Pro18, $\alpha$Arg19, $\alpha$Val20, $\alpha$Glu22, $\alpha$Trp33 in knob `a', which stabilize the high-affinity bound state (snapshots 2b,3b in Fig.~\ref{fig:fig3}). 
This flap's displacement also triggers loop I and interior region straightening, which results in formation of additional binding contacts between residues $\gamma$Lys321, $\gamma$Phe322, $\gamma$Glu323 in loop I and residues $\alpha$Glu22, $\alpha$Arg23, $\alpha$His24 in knob `a', and between residues $\gamma$Asn337, $\gamma$Cys339, $\gamma$His340, $\gamma$Ala341, $\gamma$Asn361, $\gamma$Gly362, $\gamma$Tyr363 in interior region and residues $\alpha$Gly17, $\alpha$Arg19, $\alpha$Glu22, $\alpha$Arg23, $\alpha$His24, $\alpha$Gln25 in knob `a' (Fig.~\ref{fig:figS8}d). 
The binding pocket shrinks and binding interface narrows, which results in the prolongation of bonds lifetimes.

\textbf{Role of movable flap:}
Next, we correlated the dynamics of binding contacts for the knob `a' and entire binding interface $Q$ and for the knob `a' and movable flap $Q_{MF}$. 
The results for $f =$ 30 pN (Fig.~\ref{fig:figS8}b) reveal the non-monotonic behavior of $Q$ and $Q_{MF}$ due to reversible tilting back and forth of the movable flap concomitant with its elongation and contraction. 
This results in the flap repeatedly switching off and on interactions with the $\beta$-sheet stack of B-domain. 
Additional (cryptic) binding contacts formed between the residues in movable flap and knob `a' under tension ($\gamma$Pro299--$\alpha$Val20, $\gamma$Ser300--$\alpha$Gly17, $\gamma$Ser300--$\alpha$Pro18, $\gamma$Asp301--$\alpha$Arg19, $\gamma$Lys302--$\alpha$Glu22, $\gamma$Phe303--$\alpha$Pro18, $\gamma$Phe303--$\alpha$Trp33, $\gamma$Phe304--$\alpha$Val20) are essential for the increased strength of the A:a knob-hole bond at higher forces (Fig.~\ref{fig:fig3}). 
We quantified the movable flap elongation $D_{MF}$ by calculating the distance between movable flap and the $\beta$-sheet stack of B-domain (Fig.~\ref{fig:figS8}b). 
We selected three pairs of residues in movable flap and $\beta$-sheet stack: $\gamma$Gly296--$\gamma$Ala282, $\gamma$Asp297--$\gamma$Gly283, and $\gamma$Asp298--$\gamma$Gly284. 
For each pair, we calculated $D_{MF}$ as a function of time (force) and correlated changes in $D_{MF}$ with changes in $Q$. 
The results in Fig.~\ref{fig:figS8}b show that the increase in $Q$ from 17 to 25 contacts is accompanied by $\sim$1.5-fold increase in $D_{MF}$ from 1.5 nm to 2.3 nm, which corresponds to the moveable flap translocation from the periphery to the knob `a'. 
The decrease in $Q$ from 25 to 17 contacts is correlated with the decrease in $D_{MF}$ from 2.3 to 1.3 nm, which corresponds to the movable flap tilting back (Fig.~\ref{fig:figS8}d). 
Similar results were observed for $f =$ 35 and 40 pN (data not shown). 
Therefore, the binding affinity increase and the movable flap translocation are positively correlated. 
Hence, the A:a knob-hole complex transformation to the high-affinity conformation is controlled by tension-dependent translocation of the movable flap (Fig.~\ref{fig:fig3}).

\textbf{Dynamics remodeling of A:a association interface:}
We estimated the interface width $X$ using the radius of cross-sectional area formed by connecting residues $\gamma$Asp297, $\gamma$Glu323, and $\gamma$Asn361 in the movable flap, loop I and interior region, respectively. 
The results are presented in Fig.~\ref{fig:fig4}a,b, which shows that in Pathway 1 (low-affinity bound state) $X$ fluctuates between 1.3 nm and 2.5 nm (interface wide open), whereas in Pathway 2 (high-affinity bound state) $X$ decreases from 1.3 nm to 0.8 nm (interface closes). 
Hence, the movable flap translocation also results in the hole `a' closing as well as decreasing of the interface width. 
Next, we analyzed the entire probability distributions of $X$, $P(X)$ presented in Fig.~\ref{fig:fig4}c and d. 
The profiles of $P(X)$ in Fig.~\ref{fig:fig4}c, corresponding to the trajectory of $X$ from Fig.~\ref{fig:fig4}b move towards smaller $X$, thus, displaying continuous dynamics of the A:a knob-hole binding interface closing. 
Yet, the profiles $P(X)$ in Fig.~\ref{fig:fig4}d reveal more discrete-like (two-state) dynamics of interface remodeling. 
Hence, quantitative analysis shows the evidence that hole `a' is fluctuating bottleneck with tension-dependent width of the A:a knob-hole interface, i.e. $X=X(f)$.
We were not able to discriminate between the discrete and continuous modes of evolution of $X$ due to limited number of simulation runs, which took almost two years to complete.

\subsection*{Fluctuating bottleneck theory}

\textbf{Bottleneck model of knob `a' binding pocket:} 
The main results from \textit{in vitro} and \textit{in silico} assays are the following: i) the A:a knob-hole bond first becomes stronger (increased affinity) and then becomes weaker (decreased affinity) with increasing force; ii) the force application promotes structural rearrangements resulting in binding interface remodeling; and iii) there is a competition between the dynamics of hole `a' closure and kinetics of knob `a' escape. 
To capture these observations, we propose a model which treats the binding pocket (hole `a') as a fluctuating bottleneck.\cite{ZwanzigJCP92, BarsegovJCP02, HyeonPRL14, Bicout&SzaboJCP98, Wang&WolynesCPL93} 
Pulling force affects both the dynamics of binding interface remodeling and kinetics of knob `a' escape, and so the A:a bond lifetime $\tau$ and interface width $X$ are coupled (Fig.~\ref{fig:fig4}e). 
Larger/smaller $X$ facilitates faster/slower unbinding with shorter/longer bond lifetime $\tau$. 
The A:a knob-hole bound state population $P(X,t)$ is described by the kinetic equation:
\begin{equation}\label{eq:eq1}
\frac{dP(X,t)}{dt} = -K(X)P(X,t) + \mathbf{L}(X,t)		
\end{equation}
In equation (\ref{eq:eq1}) above, the first term describes the kinetics of knob `a' escape from the bottleneck of size $X$ with rate
\begin{equation}\label{eq:eq2}
K(X) = k X^\alpha
\end{equation}
depending on shape parameter $\alpha$ (bottleneck geometry), and escape rate constant $k$. 
The rate constant is expected to increase with force, and here we use the Bell model \cite{BarsegovJPCB06, BellScience78} for $k(f) = k_0 \exp [\sigma_y f/k_B T ]$, where $k_0$ is the attempt frequency and $\sigma_y$ is the transition distance for dissociation. 
Simulations show that i) $X$ is determined by the cross-sectional area, and so we set $\alpha =$ 2 in equation (\ref{eq:eq2}); and that ii) characteristic timescale $\eta$ (milliseconds) is much shorter than bond lifetimes (seconds; Fig.~\ref{fig:fig2}), and so we set $X = \langle X \rangle$. 
In equation (\ref{eq:eq1}), operator $\mathbf{L}$ describes the dynamics of $X$:
\begin{equation}\label{eq:eq3}
\mathbf{L} =  \frac{\partial}{\partial X} \left[ \frac{k_B T}{\zeta} \frac{\partial}{\partial X} + \frac{\kappa}{\zeta} (X -\langle X \rangle) \right]	
\end{equation}
By substituting equations (\ref{eq:eq2}) and (\ref{eq:eq3}) in equation (\ref{eq:eq3}), we arrive at the Smoluchowsky equation for $P(X,t)$:
\begin{equation}\label{eq:eq4}
\frac{\partial P(X,t)}{\partial t} = \left[\frac{k_B T}{\zeta} \frac{\partial^2}{\partial X^2} + \frac{\kappa}{\zeta} (X -\langle X \rangle) \frac{\partial}{\partial X} + \frac{\kappa}{\zeta} \right] P(X,t) - k \langle X \rangle^2 P(X,t)
\end{equation}
which can be solved as described in the SI to obtain the Green's function solution given by $G(X,X_0;t)$. 
To obtain the distribution of bond lifetimes $P(t)$, we average over the initial values ($X_0$) and sum over the final values ($X$),
\begin{equation}\label{eq:eq5}
P(t) = \int_0^\infty dX \int_0^\infty dX_0 G(X,X_0;t)P(X_0)
\end{equation}
From simulations, the initial values of $X$ are sharply peaked at a fixed value $x_0$, and so $P(X_0) = \delta (X_0 - x_0)$. By substituting $P(X_0)$ and expression for $G(X,X_0;t)$ into equation (\ref{eq:eq5}) and performing the integration, we obtain:
\begin{equation}\label{eq:eq6}
P(t) = \frac{k \langle X \rangle^2}{2} \exp \left[ -k \langle X \rangle^2 t \right] \left[1 + Erf \left( \frac{\langle X \rangle}{[ 2 \pi k_B T(1 - e^{-2t/\eta})]^{1/2}} \right) \right]
\end{equation}
The average bond lifetime is given by
\begin{equation}\label{eq:eq7}
\langle \tau \rangle = \int_0^\infty t P(t) dt
\end{equation}
In the \textit{continuous version}, the binding pocket size $X$ changes continuously (Fig.~\ref{fig:fig4}c), and the average size is given by $\langle X(t) \rangle = [ X_0 e^{-t/\eta} + (x_0-f/\kappa) (1 - e^{-t)/\eta}) ] \Theta (f_c -f) + x_1 \Theta (f - f_c)$, where $\Theta(x)$ is the Heaviside step function, $X_0$ is the initial value and $\eta = \zeta/\kappa$ is the characteristic time for conformational fluctuations of the bottleneck (see SI). 
At a critical force $f \approx f_c$, $\langle X \rangle$ should reaches the minimum $x_1 = x_0 - f_c/\kappa$ (interface is closed). 
In the \textit{discrete version}, $X$ is a discrete random variable (Fig.~\ref{fig:fig4}d) interconverting between the open state $X = x_0$ and closed state $X = x_1 < x_0$ (Fig.~\ref{fig:fig4}d) with the populations $s_0 = 1/(1 + \exp [- \varepsilon_0/k_BT])$ and $s_1 = 1 - s_0$ ($\varepsilon_0$ is the energy difference). 
Pulling force favors the closed conformation by changing the state populations, and so $ \langle X(f) \rangle = x_0 s_0 (f) + x_1 s_1 (f)$ and $s_0(f) = [1 + \exp [-(\varepsilon_0 - \sigma_x f )/k_B  T)]^{-1}$, where $\sigma_x$ is the transition distance. 
The critical force is defined as force $f = f_c$ at which $s_1(f) \gg s_0(f)$ (see SI).

\textbf{A:a knob-hole bond lifetimes: insights into molecular dimensions, and nanomechanics:} 
We performed a fit of theoretical curves of $\langle \tau \rangle$ vs. $f$ calculated using equations (\ref{eq:eqS2}), (\ref{eq:eqS3}), (\ref{eq:eq6}) and (\ref{eq:eq7}) to experimental data points (Fig.~\ref{fig:fig2}b). 
The friction coefficient was set to $\zeta = 10^{-3}$ pN nm$^{-1}$s (diffusion of amino acids in water). 
The performance of continuous and discrete versions of the model compared in Fig.~\ref{fig:fig5} shows excellent agreement between theory and experiment. 
Model parameters are in Table~\ref{tab:tab1}. 
The continuous model has five parameters: interface stiffness $\kappa$, force-free interface width $x_0$, minimal interface width (hole `a' is closed) $x_1$, escape rate constant $k_0$, and transition distance for dissociation $\sigma_y$ (i.e. distance from the bound state to the transition state along dissociation path). 
The discrete model has additional parameters: transition distance from the low- to high-affinity bound state $\sigma_x$ and energy difference between these states $\varepsilon_0$ (Table~\ref{tab:tab1}). 
Parameters $\kappa$, $x_0$, and $x_1$ can be directly accessed in the simulations, which revealed the following ranges: 10-30 pN/nm for $\kappa$, 2.0-2.7 nm for $x_0$, and 0.5-0.9 nm for $x_1$. 
Hence, the continuous model provides better agreement with simulations (Table~\ref{tab:tab1}).

\section*{Discussion}

We employed the optical trap to measure the dynamic strength of single A:a knob-hole bonds -- the strongest non-covalent interactions in fibrin polymerization (Fig.~\ref{fig:fig1}). 
We used the force-clamp to profile the average bond lifetime $\langle \tau \rangle$ as a function of tensile force $f$ applied to dissociate the fibrin-fibrinogen complex. 
We found that the bond lifetimes show the biphasic catch-slip behavior, namely the average bond lifetime $\langle \tau \rangle$ first increases and then decreases with $f$ (Fig.~\ref{fig:fig2}). 
Physiological importance of catch-slip behavior can be exemplified with shear-enhanced platelet adhesion on vWF-coated surfaces where GP1b$\alpha$-vWF catch bonds promote primary hemostasis at the sites of vessel wall injury.\cite{SavageCell96, YagoJCI08, McEverARCDB10} 
Theoretical models have been proposed to explain the dynamic transition from catch-bonds to slip-bonds.\cite{BarsegovPNAS05,BarsegovJPCB06, PrezhdoACR09, ThomasCOSB09, ZhuBiorheol05, Chen&AlexanderKatzBJ11, SarangapaniJBC11} 
Yet, the structural origins of this counterintuitive behavior are not yet understood.

We showed that force signals measured in the pulling experiments represent the strength of individual knobs `a' coupled to holes `a'. 
First, we adjusted the surface densities of the reacting molecules, so that the incidence of interactions lasting $>$0.5s (i.e. representing A:a bonding) did not exceed 10\% of the total number of surface-to-surface contacts (binding attempts).
This implies that forced dissociation events corresponding to multiple A:a knob-hole bonds were highly unlikely due to sufficiently low surface density of reacting molecules (300 nm$^2$/molecule on a bead) and a small contact area (450 nm$^2$). 
Second, when the fibrin-fibrinogen interactions were measured with the force-ramp, we only observed a sharp single peak in the histograms of bond rupture forces with very rare jagged signals.\cite{LitvinovBlood05} 
The unimodal nature of distributions of bond rupture forces implies that unbinding signals represents dissociation of single knob-hole bonds. 
Multiple interactions might have occurred in some of the measurements, but infrequently, resulting in strong bead attachments lasting $>$60 s. 
These were discarded and not used in subsequent data analysis. 
Experiments with proteolytic fibrin(ogen) fragments bearing both knobs `a' and `b' (fragment desAB-E) or only knob `a' (fragment desA-E) showed catch-slip transition in both fragments, but in the absence of knob `b' the bond lifetimes were shorter and the catch-slip transition peak was less pronounced (Fig.~\ref{fig:fig2}b).
Hence, the catch-slip behavior was enhanced when both knobs `a' and `b' were exposed, which suggests positive cooperativity between knobs `a' and `b'.

We employed MD simulations of atomic structural models of the A:a knob-hole complex (Fig.~\ref{fig:fig1}g) to resolve the structural basis and to illuminate the molecular mechanism(s) of dynamic transition from catch-bonds to slip-bonds in fibrin (Methods). 
We used low damping coefficient $\gamma =$ 3.0 ps$^1$ for more efficient sampling of the conformational space \cite{FalkovichPSSA10} and to minimize the effect of fast $\nu_f = 10^3 - 10^4$ $\mu$m/s pulling speeds we had used due to very long computational time (it took $\sim$24 months to complete the computational tasks on 4 GPUs GeForce GTX780). 
Using dynamic force-ramp simulations, we were able to establish kinetic partitioning of the A:a knob-hole bond disassembly into the rapid/slow dissociation paths (Pathway 1/2) corresponding to lower/higher rupture forces (Figs.\ref{fig:fig3} and S7). 
To show that observed variation of rupture forces was not due to statistical fluctuations, we probed the dynamics of binding contacts between residues in knob `a' and hole `a' defining the binding affinity $Q$. 
For example, in Pathway 1, $Q$ decreased to zero (with some fluctuations) starting from 20 contacts (low-affinity bound state); in Pathway 2, $Q$ increased initially to 30-35 contacts (high-affinity bound state) and then decreased to zero at a higher rupture force (Fig.~\ref{fig:figS8}a; see also Fig.~\ref{fig:figS7}).

A 1.5-fold enhancement in binding affinity points to tension-induced bond stabilization through recruitment of additional binding contacts that become available for interaction with knob `a' at higher tensile forces. 
We analyzed entire maps of residue-residue binding contacts (Fig.~\ref{fig:figS8}c,d), which revealed the important role played by the movable flap. 
In the low-affinity bound state, the movable flap -- one of the three binding determinants in hole `a' -- is far away from knob `a' (snapshots 2a,3a in Fig.~\ref{fig:fig3}); therefore, the low-affinity bound state facilitates rapid detachment of knob `a' which slips easily from hole `a'. 
In the high-affinity bound state, the movable flap translocates toward and catches knob `a', forming additional binding contacts between the flap and knob `a' (snapshots 2b,3b in Fig.~\ref{fig:fig3}). 
Importantly, this transition also triggers the loop I and interior region straightening, which results in formation of additional binding contacts between knob `a' and the interior region (Fig.~\ref{fig:figS8}b). 
Therefore, in the high-affinity bound state hole `a', comprised by the movable flap, loop I, and interior region (Fig.~\ref{fig:fig1}g), shrinks and the A:a binding interface narrows, which results in labored knob `a' detachment. 
The increase in binding affinity was found to be positively correlated with the displacement of movable flap (Fig.~\ref{fig:figS8}b). 
Hence, a tension-induced increase in A:a knob-hole bond strength is entirely due to spatial rearrangement of the binding interface in hole `a'.

To better understand the biphasic catch-slip dynamic behavior of A:a knob-hole bonds, we developed new theory inspired by fluctuating bottleneck model.\cite{ZwanzigJCP92, BarsegovJCP02, HyeonPRL14, Bicout&SzaboJCP98, Wang&WolynesCPL93} 
The model treats the binding interface size $X$ as a Gaussian random variable (fluctuating bottleneck) and accounts for the tension-dependent decrease in $X$ (see equations (\ref{eq:eqS2}) and (\ref{eq:eqS3})). 
Simulations showed that $X$ is roughly equal the dimension (radius) of binding interface. 
For this reason we used a quadratic sink $K(X) \sim kX^2$ with the Bell-type dependence of $k$ on $f$ (equations (\ref{eq:eq2}), (\ref{eq:eqS5}), (\ref{eq:eqS6})).
Because we were unable to determine from simulations whether the interface interconverted between several or many conformational states, we considered a discrete (two-state) version and continuous version of the model (Fig.~\ref{fig:fig4}).
We obtained the distribution of A:a knob-hole bond lifetimes $P(t)$ (equation (\ref{eq:eq6})), which was then used to calculate $\langle \tau \rangle$ as a function of $f$. 
Excellent agreement between theoretical curves of $\langle \tau \rangle$ and experimental data points was achieved (Fig.~\ref{fig:fig5}), which enabled us to estimate parameters of the model accumulated in Table~\ref{tab:tab1}. 
Both continuous and discrete versions of the model give similar values of knob `a' escape rate $k_0 =$ 0.11-0.12 nm$^2$s$^{-1}$ and transition distance $\sigma_y =$ 0.25-0.27 nm, but the continuous version shows better agreement with estimates of other model parameters from simulations: the interface stiffness $\kappa =$ 15.7 pN/nm (theory) vs. 10-30 pN/nm (simulations), initial (force-free) interface size $x_0 =$ 2.7 nm (theory) vs. 2.0-2.7 nm (simulations), and minimal interface size $x_1 =$ 0.74 nm (theory) vs. 0.5-0.9 nm (simulations). 
Hence, our theory provides evidence for a manifold of high-affinity bound states with continuous dependence of binding affinity on $f$. 
We also calculated the profiles of $\langle \tau \rangle$ vs. $f$ but for a linear sink $K(X) \sim X$ in equation (\ref{eq:eq2}), yet, the values of model parameters showed worse agreement with simulations (see Table~\ref{tab:tabS2}).

We used values of model parameters for the continuous fluctuating bottleneck model with quadratic sink (Table~\ref{tab:tab1}), which demonstrated the best agreement with the simulations, to calculate the distributions of bond lifetimes $P(t)$.
Theoretical curves of $P(t)$ are compared with experimental histograms in Fig.~\ref{fig:figS9}, which shows that experiment and theory agree very well. 
Using the values of model parameters (Tables \ref{tab:tab1} and \ref{tab:tabS2}) we compared the prediction of all four models (continuous vs. discrete version with quadratic vs. linear sink) for the critical force $f_c$ of transition from the catch-to-slip regime of dissociation of A:a knob-hole bond. We found that for the continuous version $f_c =$ 30.7 pN (quadratic sink) and 30 pN (linear sink), whereas for the discrete version $f_c =$ 44.6 pN (quadratic sink) and 40.1 pN (linear sink). 
Hence, the continuous model provides a better prediction for the 30-35 pN critical force, which corresponds to the 4-5 s maximum bond lifetime (Fig.~\ref{fig:fig5}).

To conclude, we demonstrated experimentally, resolved computationally, and modeled theoretically the catch-slip dynamic transition in A:a knob-hole bonds in fibrin. The movable flap plays an important role of a tension-dependent molecular switch, which triggers the crossover from the catch regime to the slip regime of bond dissociation. 
In the catch regime, the binding affinity of hole `a' progressively grows with force owing to mechanical interface remodeling and formation of additional binding contacts, which results in bond strengthening. This trend continues until the size of binding pocket becomes comparable with the molecular dimension of knob `a' at a critical force $f = f_c$, at which point the slip regime sets in. 
In the slip regime, the binding affinity of hole `a' gradually decreases with the increasing force $f > f_c$ due to disruption of binding contacts, which weakens the bond. 
The fluctuating bottleneck theory can be used to model biomolecular complexes displaying rich complex dynamics of interface remodeling.

(Patho)physiologically the catch-bond behavior is equivalent to shear-enhanced strengthening of A:a knob-hole bonds that might favor fibrin polymerization in blood flow and might prevent breakup and damage of stressed clots in vasculature. 
This effect is especially important at the early stages of fibrin formation in blood flow when the incipient clot is small and the hydrodynamic shear stress is low, so that the tensile forces are in the range that strengthens the knob-hole interactions. 
This is an entirely new aspect of fibrin nanomechanics that needs further investigation. 
The novel and unexplored mechano-chemical aspects of fibrin polymerization addressed in this study for the first time will advance our understanding of blood clotting and will provide a firm foundation for the development of new approaches to control and modulate this process.

\section*{Online Methods}

\textbf{Optical trap-based model system:} 
Our model system for probing the bimolecular interactions is based on an optical trap that uses a focused laser beam to generate the pico-Newton mechanical force to hold and move microscopic particles, such as micron-size polystyrene beads.\cite{LitvinovPNAS02,LitvinovBlood05, LitvinovBJ05, LitvinovBJ11, LitvinovJBC12, LitvinovJBC16}
A custom-built optical trap previously described in detail \cite{LitvinovBJ11} was used to measure the mechanical strength of individual bi-molecular protein-protein complexes under a constant tensile force. 
The core of the laser tweezers system is a AxioObserver Z1 inverted microscope and a 100x 1.3NA Fluor lens combined with a FCBar Nd:YAG laser (=1,064 nm) with 4W power in continuous TEM-00 mode. 
A computer-operated two-dimensional acousto-optical deflector (AOD) was used to control the trap position. 
The force exerted by the trap on the bead displaced by an amount $\Delta x$ was measured with a quadrant detector and the trap position was corrected with an electronic feedback loop to keep the force constant. 
This system enabled control of the duration of compressive contact between interacting surfaces $T$, the magnitude of compressive force $f_c$ and the magnitude of the tensile force $f = k_{opt} \Delta x$ ($k_{opt}$ is the optical trap stiffness). 
The measured quantity was the time needed to separate the interacting surface-attached proteins (bond lifetime) $\tau$. 
All experiments were conducted with the average trap stiffness of $k_{opt} =$ 0.10$\pm$0.02 pN/nm. 
Force calibration and trap stiffness were routinely confirmed by the Stokes' force method. 
LabVIEW$\circledR$ software was used to control and record laser beam deflection, to move the piezoelectric stage, and to analyze data off-line.

\textbf{Surfaces and proteins:}
A single fibrinogen- or fragment D-coated bead was trapped and repeatedly brought into contact with a fibrin- or fragment E-coated pedestal. When fibrinogen or fragment D (with holes `a' and `b') on the bead attached to monomeric fibrin or activated fragment E (with both knobs `a' and `b' or knobs `a' only) on the pedestal, the trap exerted a constant force $f = k_{opt} \Delta x$ to trigger the complex dissociation.\cite{LitvinovBlood05} 
Purified human fibrinogen or its fragment E (both from HYPHEN BioMed, France) were bound covalently to spherical silica pedestals 5 m in diameter anchored to the bottom of a chamber. 
Pedestals coated with a thin layer of polyacrylamide were activated with 10\% glutaraldehyde, after which the proteins were immobilized overnight at 4 $^\circ$C from 1 mg/ml solution in 20 mM HEPES pH 7.4 containing 150 mM NaCl and 3 mM CaCl2.
After washing off the non-covalently adsorbed protein, 2 mg/ml bovine serum albumin (BSA) in 0.055 M borate buffer pH 8.5 was added as a blocker. 
To form fibrin-coated pedestals, the immobilized fibrinogen was treated with human $\alpha$-thrombin (Enzyme Research Laboratories, South Bend, IN) (1 U/ml, 37 $^\circ$C, 1 hr) followed by washing of the chambers with 20 volumes of 100 mM HEPES pH 7.4 containing 150 mM NaCl, 3 mM CaCl2, 2 mg/ml BSA, and 0.1\% (v/v) Triton X-100 before the measurements. 
To form pedestals coated with fibrin fragment E bearing knobs `a' (fragment desA-E) or knobs `a' and `b' (fragment desAB-E), the immobilized fibrinogen fragment E with uncleaved fibrinopeptides was treated with batroxobin (Batroxobin moojeni, CenterChem, Stamford, CT) (1 BU/ml, 37 $^\circ$C, 1 hr) or human $\alpha$-thrombin (1 U/ml, 37 $^\circ$C, 1 hr), respectively, followed by washing of the chambers. 
1 unit of batroxobin activity (BU) was calibrated to be equal to 1 unit of thrombin activity (U) in terms of the rate of fibrinopeptide A release. 
Fibrinogen or fragment D (HYPHEN BioMed, France) were bound covalently to carboxylate-modified 1.75-m latex beads (Bangs Laboratories, Carmel, IN) activated by N-(3-dimethylaminopropyl)-N'-ethylcarbodiimide hydrochloride. 
The immobilization step lasted 15 min at 4 $^\circ$C in 0.055 M borate buffer pH 8.5 containing 150 mM NaCl and 3 mM CaCl2. 
BSA was used as a blocker. When immobilized from 20 g/ml solution containing 100\% of fibrinogen labeled with I$^{125}$, the surface density of I$^{125}$-fibrinogen was determined to be about (11$\pm$2)$\times$10$^{-9}$ g/m$^2$, which approaches the point of surface binding saturation. 
Both fibrinogen and fragment D were used with the same solution concentration (20 g/ml).

\textbf{Measurements of protein-protein interactions:} 
Experiments were performed at room temperature in 100 mM HEPES pH 7.4 containing 150 mM NaCl, 3 mM CaCl2 with 2 mg/ml BSA and 0.1\% (v/v) Triton X-100 added to reduce non-specific interactions. 
1 l of the fibrinogen- or fragment D-coated bead suspension (10$^7$ beads/ml) was added to 50 l of the working buffer and flowed into a chamber containing pedestals with immobilized fibrin or fragment E on their surface. 
After the chamber was placed on the microscope stage, a single bead was trapped and the stage moved manually to bring a pedestal within 1-2 microns of the trapped bead. 
After starting the bead oscillation, the separation of the pedestal and the bead was then reduced until they touched each other repeatedly with a compressive force $f_c =$ 20-30 pN and contact duration $T =$ 0.5 s. 
The constant pulling force was varied from $f =$ 5 to 60 pN. 
Trap displacement signals were recorded at 2000 scans per second and a bond lifetime was measured for each pedestal-bead touching event. 
Several tens of pedestal-bead pairs were analyzed for each set of conditions. 
The binding-unbinding events from individual files were summarized; the total number of bond lifetime values recorded for each set of experimental conditions varied from $\sim$3,000 to $\sim$4,500. 
The bond lifetimes $<$0.5 s represented non-specific interactions and were not susceptible for specific inhibition. 
These short bond lifetimes were not included into data analysis and modeling.

\textbf{All-atomic structural model of A:a knob-hole complex:}
The N-terminal motif Gly-Pro-Arg (GPR) in $\alpha$ chain is the main functional sequence of the knob `a', and it is complementary to the hole `a' in $\gamma$-nodule of adjacent fibrin molecule. 
The N-terminal $\beta$ chain motif Gly-His-Arg-Pro (GHRP) is a major part of the knob `b' that binds to the hole `b' located in $\beta$-nodule of adjacent fibrin.\cite{MedvedJTH09} 
The available atomic structures (PDB entry 1FZA, 1FZB, 1FZC)\cite{EverseBiochemistry98} contain only the D:D interface, i.e. the $\beta$- and $\gamma$-nodules of two cross-linked fibrin monomers with bound knob-mimetic peptides GPRP and GHRP, which occupy the corresponding binding pockets in $\gamma$- and $\beta$-nodules, respectively (Fig.~\ref{fig:fig1}). 
We used the PDB data to reconstruct a physiologically relevant model of the A:a complex comprising the $\gamma$-nodule bound with the N-terminal end of the $\alpha$-chain of adjacent fibrin monomer. 
We utilized the complete atomic structure of a short fibrin oligomer.\cite{ZhmurovStructure16} 
The model contains the N-terminal ends of $\alpha$-chains (with knobs `a') and $\beta$-chains (with knobs `b') bound to the complementary sites in the $\gamma$-nodule (with holes `a') and $\beta$-nodules (with holes `b') as shown in Fig.~\ref{fig:fig1}e. 
All \textit{in silico} models were constructed using CHARMM.\cite{Charmm09}
This structure was equilibrated for $\sim$100 ns using the MD simulations in implicit solvent. 
The equilibrated structure of the fibrin trimer was truncated to separate the central nodule of one trimer strand with two exposed knobs `a' bound to the complementary holes `a' of the $\gamma$-nodules in the other strand (Fig.~\ref{fig:fig1}f).
To mimic experimental conditions of tensile force application, we constrained the C$_\alpha$-atoms of $\gamma$Lys159 of two $\gamma$-nodules and pulled at the entire central domain. 
This resulted in a partial extension of $\alpha$ chains. 
We separated the $\gamma$-nodule (residues $\gamma$Lys140--$\gamma$Val411) with attached N-terminal part of $\alpha$ chain (residues $\alpha$Gly17--$\alpha$Cys36).
The obtained structural model of A:a knob-hole complex is shown in Fig.~\ref{fig:fig1}g. 
The $\gamma$-nodule (with hole `a') consists of three globular domains called A-domain ($\gamma$Val143--$\gamma$Trp191), B-domain ($\gamma$Thr192--$\gamma$Ala286 and $\gamma$Lys380--$\gamma$Leu392), and P-domain ($\gamma$Gly287--$\gamma$Met379). 
The latter is a knob `a'-binding domain,\cite{YeeStructure97} while the B-domain has a 3-stranded $\beta$-sheet stack (regulatory domain; residues $\gamma$Ile242--$\gamma$Gly283) located near the P-domain (binding domain). 
The $\gamma$-nodule has three binding determinants: loop I (region I; $\gamma$Trp315--Trp330), interior region (region II; $\gamma$Trp335--Asn365), and moveable flap (region III; $\gamma$Phe295--Thr305) displayed in Fig.~\ref{fig:fig1}g,\cite{LaudanoPNAS78, KostelanskyBiochem02} which define the strength of A:a knob-hole bond.\cite{KononovaJBC13}

\textbf{Dynamic force measurement of the A:a knob-hole interactions:}
We employed the all-atom MD simulations using Solvent Accessible Surface Area (SASA) model of implicit solvation with CHARMM19 unified hydrogen force-field \cite{FerraraProteins02, ZhmurovJACS12, KononovaBiochem17} implemented on a GPU. 
We used a lower damping coefficient $\gamma =$ 3.0 ps$^{-1}$ (vs. $\gamma =$ 50 ps$^{-1}$ for ambient water at 300K) for more efficient sampling of the conformational space.\cite{FalkovichPSSA10} 
In the force-ramp measurements, we used time-dependent force $f(t) = k_{opt}(\nu_f t - \Delta x)$, where $\nu_f$ is the virtual optical trap velocity, $k_{opt}$ is the spring constant, and $\Delta x$ is the displacement of a pulled residue. 
We constrained the C$_\alpha$-atoms of $\gamma$Lys159 and pulled the C$_\alpha$-atom of $\alpha$Cys36 in the direction perpendicular to the A:a binding interface (Fig.~\ref{fig:fig1}g). 
We generated a total of 20 simulation runs with $\nu_f = 10^3$, $10^4$ $\mu$m/s and $k_{opt} =$ 100 pN/nm. 
In the force-clamp measurements, we used constant tensile force $\mathbf{f} = f \cdot \mathbf{n}$ with force magnitude $f =$ 30, 35, 40, 50, 80 pN (Fig.~\ref{fig:fig1}g). 
We performed 3 simulation runs (a total of 5 $\mu$s) for each force value.

\section*{Acknowledgements}

The authors thank Andrey Mekler for technical assistance. 
This work was supported by NSF (grant DMR1505662 to JWW and VB), American Heart Association (grant-in-aid 13GRNT16960013 to VB and JWW), Russian Foundation for Basic Research (grant 15-37-21027, 15-01-06721 to AZ and grant 14-04-32066 to OK) and the Program for Competitive Growth at Kazan Federal University.

\section*{Author contributions statement}

R.I.L. -- designed and performed experiment, wrote the manuscript; 
O.K. -- performed simulations and analyzed the results, developed the model, wrote the manuscript; 
F.M. -- performed model fitting; 
A.Z. -- analyzed results of simulations; 
K.A.M. -- analyzed the results of simulations, wrote the manuscript; 
J.W.W. -- designed experiment, wrote manuscript; 
V.B. -- designed simulations and model, wrote the manuscript.

\section*{Competing financial interests} 
The authors declare no competing financial interests.

\bibliography{catchslip,fibrin,ours,md}

\newpage
\begin{table}[ht]
\caption{\label{tab:tab1} Model parameters for the A:a knob-hole bottleneck: interface stiffness $\kappa$, initial (maximal) and minimal interface width $x_0$ and $x_1$, escape rate constant $k_0$, transition distance for dissociation $\sigma_y$, and transition distance for conformational transition state $\sigma_x$ and energy difference between two states $\varepsilon_0$ (for discrete two-state version). 
The parameter values were obtained by fitting theoretical curves of $\langle \tau(f) \rangle$ (equation (\ref{eq:eq7}) to the experimental bond lifetimes (Fig.~\ref{fig:fig5}).}
\centering
\begin{tabular}{|c|c|c|c|c|c|c|c|}
\hline
Model & $\kappa$, pN/nm & $x_0$, nm & $x_1$, nm & $k_0$, nm$^2$s$^{-1}$ & $\sigma_y$, nm & $\sigma_x$, nm & $\varepsilon_0$, kcal/mol \\
\hline
Continuous & 15.7 & 2.7 & 0.74 & 0.11 & 0.25 & - & - \\
\hline
Discrete & 100.0 & 4.3 & 1.6 & 0.12 & 0.27 & 0.55 & 0.8 \\
\hline
\end{tabular}
\end{table}

\include{figures}

\include{supplementary}

\end{document}

%% file: figures.tex
\newpage
\begin{figure}[ht]
\centering
\includegraphics[width=0.8\linewidth]{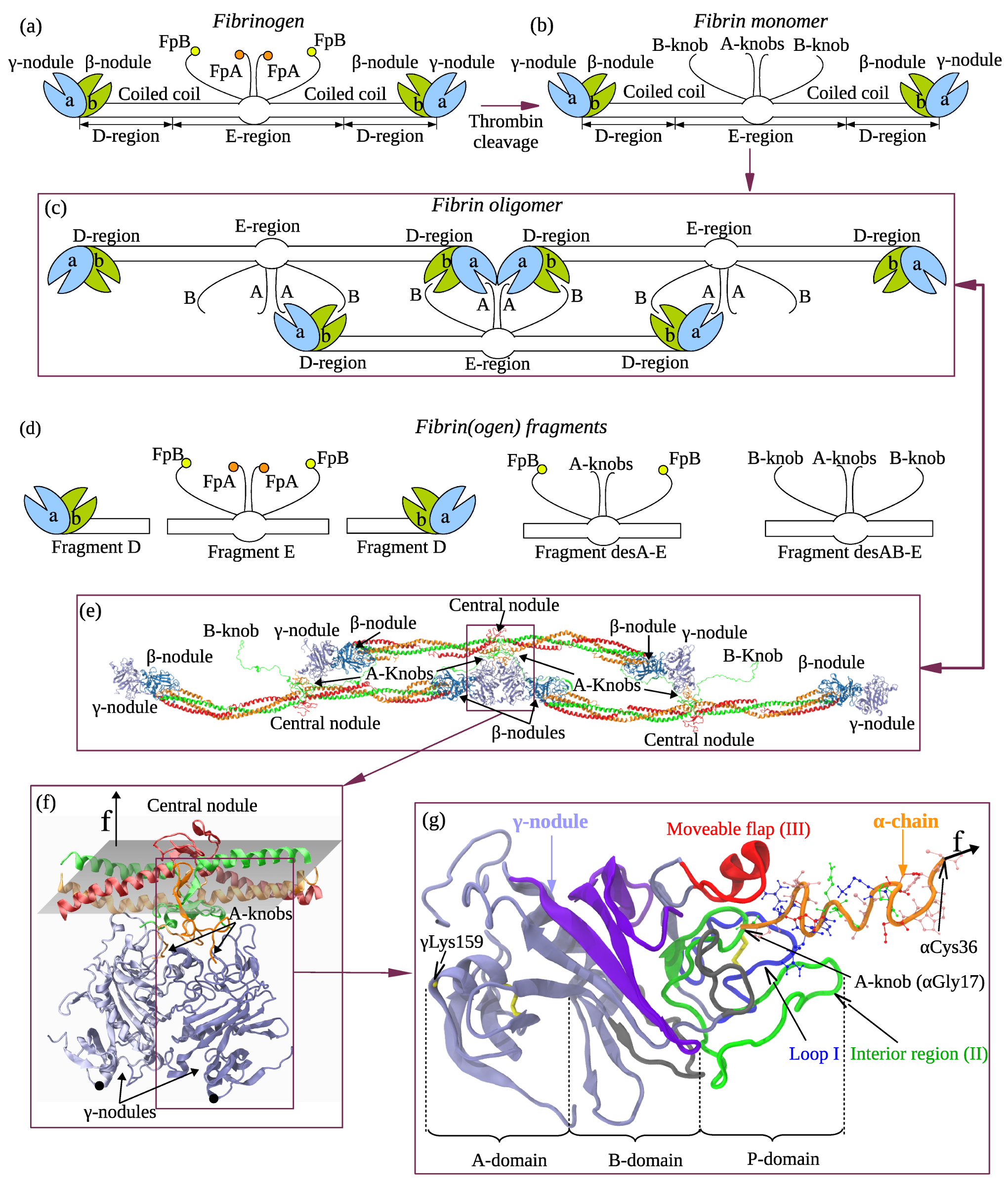}
\caption{
\textbf{Major steps and main structural determinants in fibrin polymerization.} 
Cleavage by thrombin of the N-terminal parts of the $\alpha$-chain - fibrinopeptides A and B (FpA and FpB) in fibrinogen (panel a) converts it into fibrin with exposed knobs `A' and `B' (panel b), which bind to the complemental hole `a' in the $\gamma$-nodules (blue) and `b' in the $\beta$-nodule (green), facilitating fibrin oligomerization (panel c). 
Panel d shows the fibrin fragments utilized in dynamic force spectroscopy experiments \textit{in vitro} and \textit{in silico}. 
Panel e: all-atom model of the short fibrin oligomer \cite{ZhmurovStructure16} with $\alpha$-, $\beta$- and $\gamma$-chains colored in orange, green and red, respectively. 
The $\beta$- and $\gamma$-nodules are highlighted in light blue and light grey, respectively. 
Panel f: truncated portion of fibrin oligomer (with $\gamma$-nodules and central nodule) used in simulations. 
The pulling force is applied through a virtual plane, crossing the central nodule in the middle (C$_\alpha$-atoms of $\gamma$Lys159 in the $\gamma$-nodule were constrained). 
Panel g: all-atom model of A:a knob-hole complex containing $\gamma$-nodule with hole `a' (grey blue) and the N-terminal part of $\alpha$-chain with knob `A' (orange). 
Hole `a' contains domains A, B, and P; three-stranded $\beta$-sheets stack in B-domain is shown in purple. 
Also shown are the major binding determinants: loop I (blue), interior region (green), and moveable flap (red) all interacting with GPR-motif-containing knob `A'.\cite{KononovaJBC13} 
In pulling simulations, tensile force was applied to the C$_\alpha$-atom of $\alpha$Cys36 (C$_\alpha$-atom of $\gamma$Lys159 was constrained).
}
\label{fig:fig1}
\end{figure}

\newpage
\begin{figure}[ht]
\centering
\includegraphics[width=0.7\linewidth]{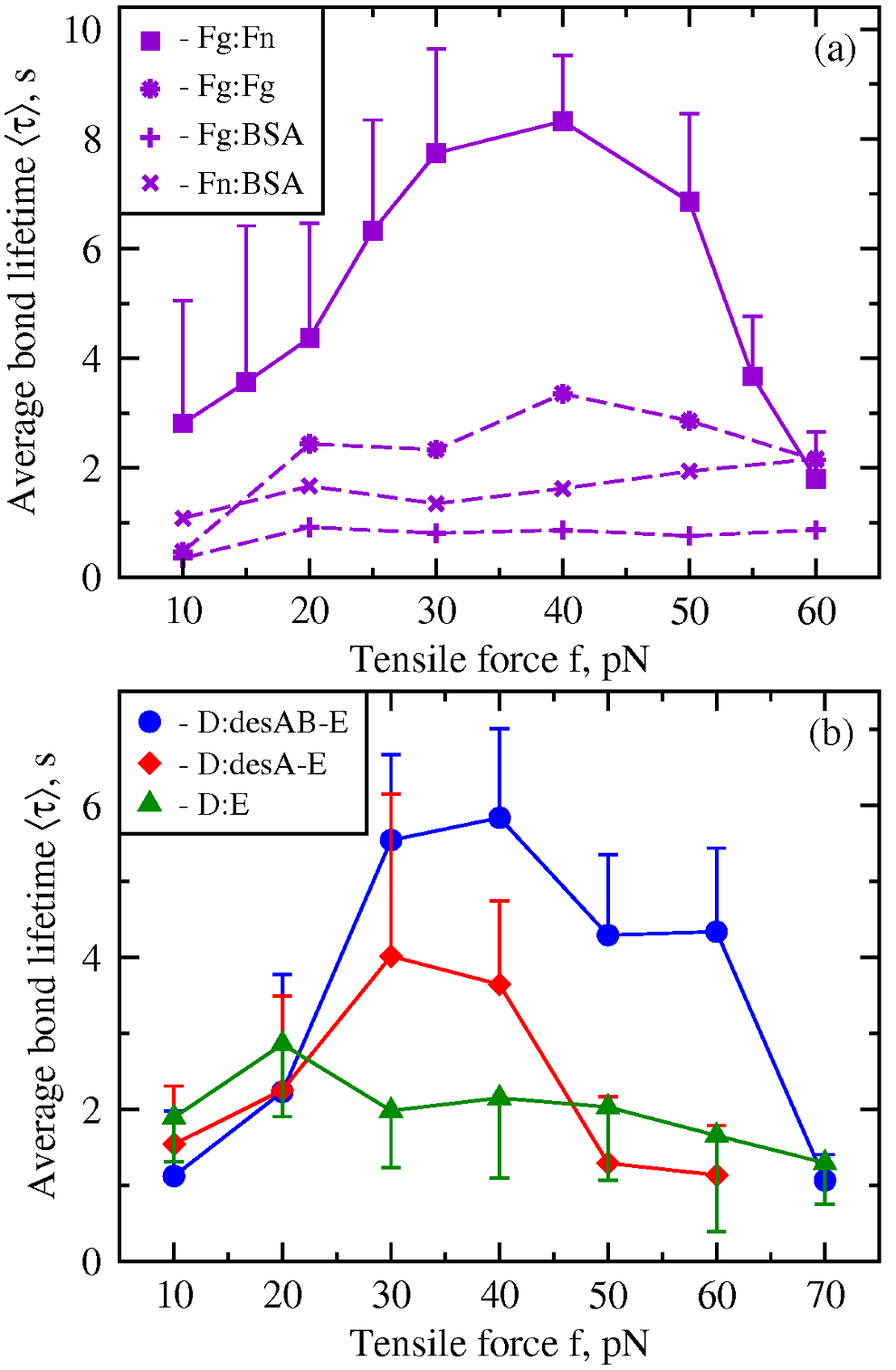}
\caption{
\textbf{Kinetics of forced dissociation of the A:a knob-hole bonds.}
Panel a: Plots of the average bond lifetime $\langle \tau \rangle$ as a function of constant tensile force $f$ for fibrin-fibrinogen (Fn:Fg) interactions in comparison with the data for control protein pairs both lacking knobs `a’ (Fg:Fg) or one of the interacting proteins (BSA), or lacking both knobs `a’ and holes `a’ (Fg:BSA and Fn:BSA). 
Panel b: Plots of $\langle \tau \rangle >$ 0.5 s as a function of $f$ for fibrinogen fragment D (with holes `a’ and `b’) interacting with fragments desAB-E (bearing knobs `a’ and `b’) or desA-E (with knobs `a’ only), or E (with no knobs).
}
\label{fig:fig2}
\end{figure}

\newpage
\begin{figure}[ht]
\centering
\includegraphics[width=0.8\linewidth]{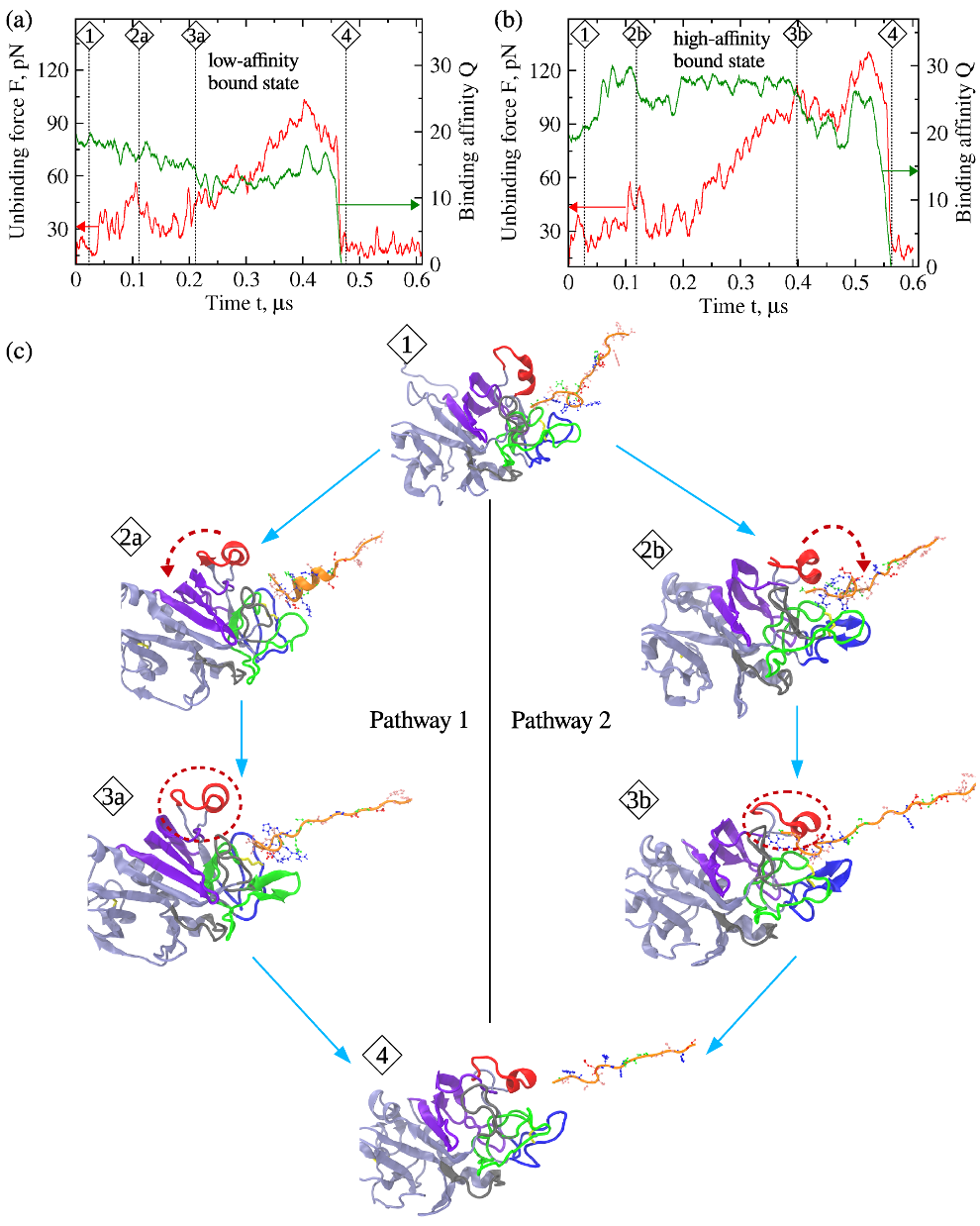}
\caption{
\textbf{Dynamic interface remodeling and A:a knob-hole bond dissociation from the low-affinity (Pathway 1) and high-affinity (Pathway 2) bound states.}
Shown are profiles of unbinding force $F$ and binding affinity $Q$ as functions of time $t$ from force-ramp simulations ($\nu_f = 10^4$ $\mu$m/s and $k_{opt} =$ 100 pN/nm). 
Pathway 1/2 is characterized by lower/higher values of $F$ and monotonic/non-monotonic evolution of $Q$. 
Structural snapshots numbered 1-4 correspond to the equally numbered regions in the curves of $F$ vs. $t$ (red) and $Q$ vs. $t$ (green) displaying the progress of dissociation from the initial bound state (structure 1) to the fully dissociated state (structure 4) along each pathways. 
Low-affinity bound state: movable flap interacts with the $\beta$-sheet stack of B-domain (red arrow and dashed circle in structures 2a and 3a). 
High-affinity bound state: movable flap moves from the periphery toward knob `a’ offering additional binding contacts; loop I and interior region straighten up, which results in the interface closing (red arrow and dashed circle in structures 2b and 3b).
}
\label{fig:fig3}
\end{figure}

\newpage
\begin{figure}[ht]
\centering
\includegraphics[width=\linewidth]{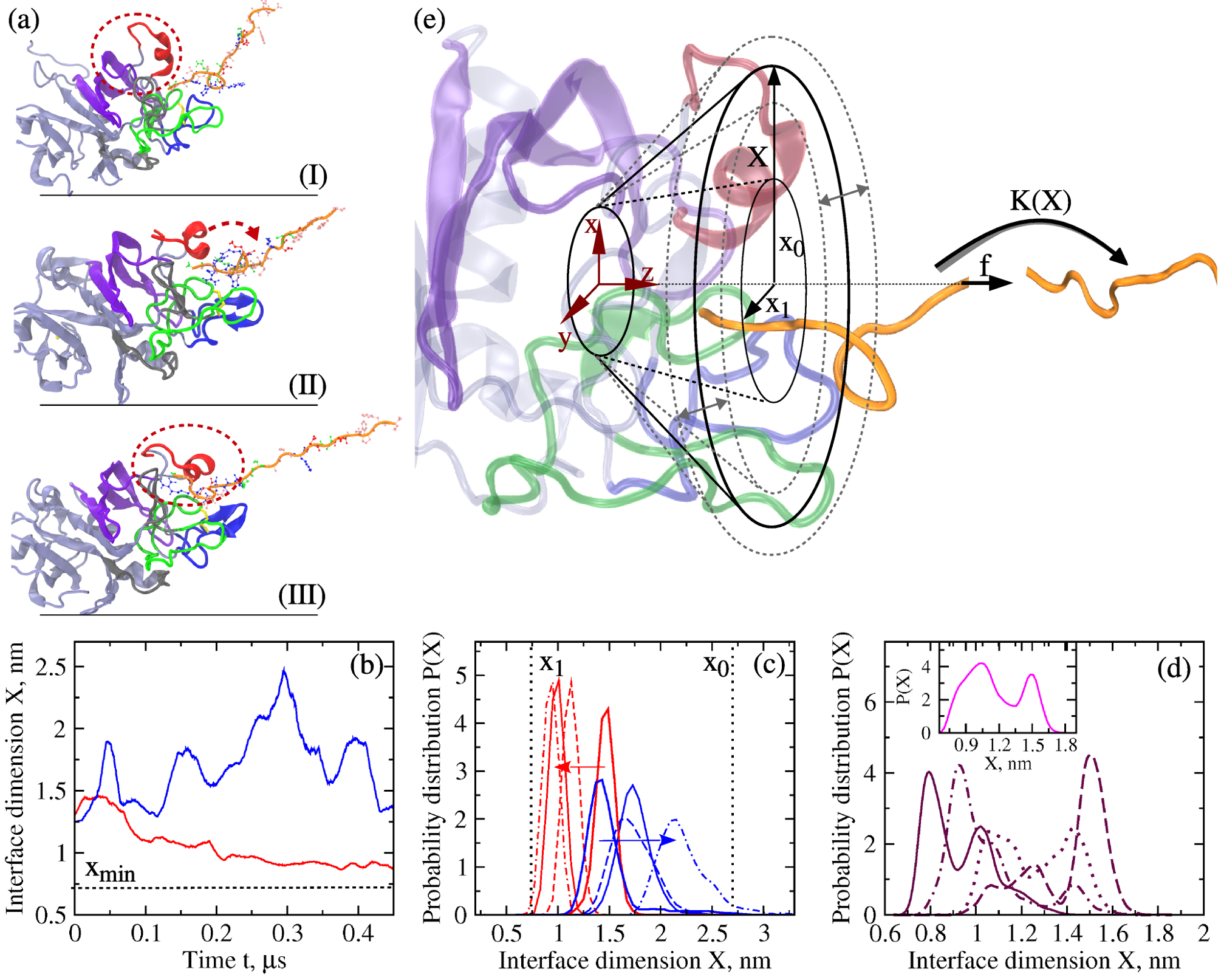}
\caption{
\textbf{Fluctuating bottleneck model of forced dissociation of A:a knob-hole bond.}
Panel a: structural snapshots I-III, showing gradual interface closing. 
Panel b: evolution of the interface width $X$ for two representative simulations resulted in bond dissociation from the low-affinity bound state (blue curve) and high-affinity bound state (red curve). 
Panel c: The probability distributions $P(X)$ of interface width $X$ for two simulations (from panel b) showing continuous evolution of $X$. 
Panel d: $P(X)$ for simulations showing discrete (two-state) evolution of $X$ from the low- to high-affinity bound states centered around $x_0$ to $x_1$, respectively (the inset shows $P(X)$ averaged over all simulation runs showing discrete evolution of $X$). 
Panel e: binding packet (hole `a’) modeled as a fluctuating bottleneck of width $X$ decreasing from the initial value $x_0$ to final value $x_1$.
}
\label{fig:fig4}
\end{figure}

\newpage
\begin{figure}[ht]
\centering
\includegraphics[width=0.7\linewidth]{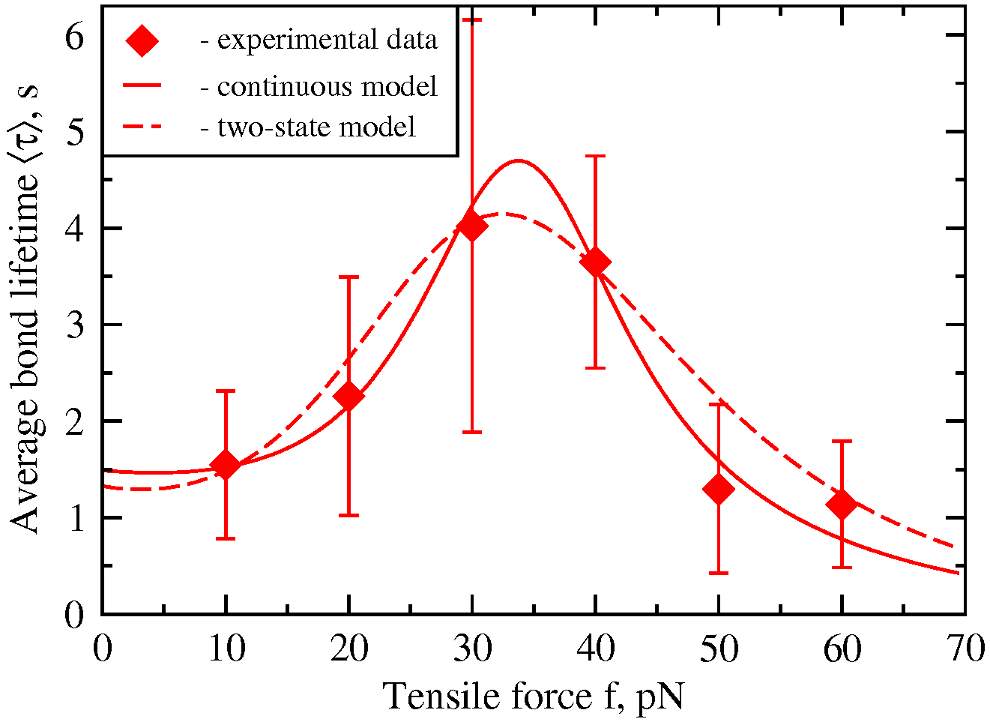}
\caption{
\textbf{Catch-slip dual force response of A:a knob-hole bond.}
Shown is the non-monotonic behavior of the average bond lifetime $\langle \tau \rangle$ with the tensile force $f$. 
The diamonds are experimental data points from Fig.~\ref{fig:fig2}b. 
The curves correspond to the fits of continuous version (equations (\ref{eq:eqS2}), (\ref{eq:eq6}), (\ref{eq:eq7})) and discrete two-state version (equations (\ref{eq:eqS3}), (\ref{eq:eq6}), (\ref{eq:eq7})) of fluctuating bottleneck model.
The initial growth of $\langle \tau \rangle$ at low forces $f < f_c \approx$ 30-35 pN is followed by the decay to zero at higher forces $f > f_c$, which marks the transition from the catch-regime to the slip-regime of the A:a knob-hole bond dissociation.
}
\label{fig:fig5}
\end{figure}

%% file: supplementary.tex
%\section*{Supplementary Information}
\renewcommand{\appendixpagename}{Supplementary Information}
\setcounter{equation}{0}
\renewcommand{\theequation}{S\arabic{equation}}
\setcounter{figure}{0}
\renewcommand{\thefigure}{S\arabic{figure}}
\setcounter{table}{0}
\renewcommand{\thetable}{S\arabic{table}}
\appendix
\appendixpage

\textbf{Fluctuating bottleneck model:}
In the continuous version of the model, the binding pocket size $X$ is a continuous random variable (Fig.~\ref{fig:fig4}e). 
Evolution of $X$ is governed by the Langevin equation: $\zeta \dot{X}(t) = -\partial U/ \partial X + R(t)$, where $\zeta$ is the friction coefficient, and $R(t)$ is Gaussian random force with the average $\langle R(t) \rangle = 0$ and standard deviation $\langle R(t)R(t') \rangle = 2\zeta k_B T \delta(t - t')$  ($k_B T$ -- temperature). 
The distributions of $X$ are Gaussian-like (Fig.~\ref{fig:fig4}c).
Hence, for $f =$ 0, $X$ evolves on a harmonic potential $U = \kappa (X - x_0)^2$ with stiffness $\kappa$ and equilibrium width $x_0$.
When $0 < f < f_c$ ($f_c$ is the critical force (see below)), $X$ decreases (Fig.~\ref{fig:fig4}b) approaching new equilibrium $x_0 - f/\kappa$, and the deterministic force is $\partial U/ \partial X = \kappa (X - x_0) + f = \kappa [X - (x_0 - f/\kappa)]$. 
By substituting this expression into Langevin equation and solving this equation, we obtain the average size $\langle X \rangle$ and fluctuations $\langle \Delta X \rangle^2$:
\begin{equation}\label{eq:eqS1}
\langle X(t) \rangle =  X_0 e^{-t/\eta} + (x_0-f/\kappa) (1 - e^{-t)/\eta}) \text{ and } \langle \Delta X \rangle^2 = \frac{k_B T}{\zeta}(1 - e^{-t)/\eta})
\end{equation}
where $X_0$ is the initial value and $\eta = \zeta/\kappa$ is the characteristic time for conformational fluctuations of the binding pocket (bottleneck). At a critical force $f \approx f_c$,  $\langle X(t) \rangle$ should reaches the minimum $x_1 = x_0 - f_c/\kappa$, which can be expressed as
\begin{equation}\label{eq:eqS2}
\langle X(t) \rangle = [ X_0 e^{-t/\eta} + (x_0-f/\kappa) (1 - e^{-t)/\eta}) ] \Theta (f_c -f) + x_1 \Theta (f - f_c)
\end{equation}
where $\Theta(x)$ is the Heaviside step function. 
In the \textit{discrete version} of the model, $X$ is a discrete random variable interconverting between the open state $X = x_0$ and closed state $X = x_1 < x_0$ (Fig.~\ref{fig:fig4}d) with the populations $s_0 = [1 + \exp(-\varepsilon_0/k_BT)]^{-1}$ and $s_1 = 1 - s_0$ ($\varepsilon_0$ is the energy difference). 
Pulling force application favors the closed conformation by changing the state populations, in which case $\langle X \rangle$ and $s_{0,1}$ become force-dependent,
\begin{equation}\label{eq:eqS3}
\langle X(f) \rangle = x_0 s_0(f) + x_1 s_1(f) \text{ and } s_0 = (1 + \exp [- (\varepsilon_0 - \sigma_x)/k_BT])^{-1}
\end{equation}
where $\sigma_x$ is the transition distance. 
The critical force is defined as force $f = f_c$ at which $s_1(f) \gg s_0(f)$.

Pulling force affects both the dynamics of binding interface remodeling and kinetics of knob `A' escape, and so the A:a bond lifetime $\tau$ and interface width size $X$ are coupled (Fig.~\ref{fig:fig4}e).
Larger/smaller $X$ facilitates faster/slower unbinding with shorter/longer bond lifetime $\tau$. 
The A:a knob-hole bound state population $P(X,t)$ is described by the kinetic equation:
\begin{equation}\label{eq:eqS4}
\frac{dP(X,t)}{dt} = -K(X)P(X,t) + \mathbf{L}P(X,t)
\end{equation}
In equation (\ref{eq:eqS4}) above, the first term describes the kinetics of knob `A' escape from the bottleneck of size $X$ with rate
\begin{equation}\label{eq:eqS5}
K(X) = k X^\alpha
\end{equation}
depending on shape parameter $\alpha$ (bottleneck geometry), and escape rate constant $k$ (when $K =$ 0; knob `A' is tightly locked in hole `a'). 
The rate constant is expected to increase with force, and here we use the Bell model:\cite{BarsegovPNAS05, BarsegovJPCB06, BellScience78}
\begin{equation}\label{eq:eqS6}
k(f) = k_0 \exp [-\sigma_y f/k_B T]
\end{equation}
where $k_0$ is the attempt frequency and $\sigma_y$ is the transition distance for dissociation. 
Simulations show that i) $X$ is determined by the cross-sectional area, and so we set $\alpha =$ 2 in equation (\ref{eq:eqS5}); and that ii) characteristic timescale $\eta$ (milliseconds) is much shorter than bond lifetimes (seconds; Fig.~\ref{fig:fig2}), and so we set $X = \langle X \rangle$ in equation (\ref{eq:eqS5}). 
In equation (\ref{eq:eqS4}), operator $\mathbf{L}$ describing the dynamics of $X$:
\begin{equation}\label{eq:eqS7}
\mathbf{L} =  \frac{\partial}{\partial X} \left[ \frac{k_B T}{\zeta} \frac{\partial}{\partial X} + \frac{\kappa}{\zeta} (X -\langle X \rangle) \right]	
\end{equation}
By substituting equations (\ref{eq:eqS6}) and (\ref{eq:eqS7}) into equation (\ref{eq:eqS4}), we arrive at the Smoluchowsky equation for $P(X,t)$:
\begin{equation}\label{eq:eqS8}
\frac{\partial P(X,t)}{\partial t} = \left[ \frac{k_B T}{\zeta} \frac{\partial^2}{\partial X^2} + \frac{\kappa}{\zeta} (X - \langle X \rangle) \frac{\partial}{\partial X}  + \frac{\kappa}{\zeta} \right] P(X,t) - k \langle X \rangle^2 P(X,t)
\end{equation}
which can be solved as described below. The Green’s function solution is given by
\begin{equation}\label{eq:eqS9}
G(X, X_0; t) = \left[ \frac{2 \pi k_B T}{\kappa}  (1 - e^{-2t/\eta})\right]^{-1/2} \exp \left[ -\frac{\kappa (X - \langle X \rangle)^2}{2 k_B T (1 - e^{-2t/\eta})} \right] k \langle X \rangle^2 \exp \left[-k \langle X \rangle^2 t  \right] 
\end{equation}
To obtain the expression for $P(t)$, we need average over the initial values ($X_0$) and sum over the final values ($X$),
\begin{equation}\label{eq:eqS10}
P(t) = \int_0^\infty dX \int_0^\infty dX_0 G(X, X_0; t) P(X_0)
\end{equation}
From simulations, the initial values of $X$ are sharply peaked fixed value $x_0$, and so we use the Dirac delta function, $P(X_0) = \delta (X_0 - x_0)$. 
By substituting $P(X_0)$ and equation (\ref{eq:eqS9}) into equation (\ref{eq:eqS10}) and performing the integration, we obtain the distribution of bond lifetimes:
\begin{equation}\label{eq:eqS11}
P(t) = \frac{k \langle X \rangle ^2}{2} \exp [-k \langle X \rangle^2 t ] \cdot \left[1 + Erf \left( \frac{\langle X \rangle}{[2 \pi k_B T/\kappa (1 - e^{-2t/\eta})]^{1/2}} \right) \right]
\end{equation}
In equation (\ref{eq:eqS11}), the average size $\langle X \rangle$ is given by equations (\ref{eq:eqS2}) and (\ref{eq:eqS3}) for the continuous and discrete versions, respectively. 
The average bond lifetime is given by
\begin{equation}\label{eq:eqS12}
\langle \tau \rangle = \int_0^\infty tP(t)dt
\end{equation}

\textbf{Derivation of equation (\ref{eq:eqS9}):}
Equation (\ref{eq:eqS8}) in the text above can be solved by using the Zwanzig's ansatz:
\begin{equation}\label{eq:eqS13}
P(t) = \left( \frac{2 \pi}{\kappa} \right)^{-1/2} \exp \left[ -\frac{(X - \langle X \rangle)^2}{2} \mu(t) - \nu(t) \right]
\end{equation}
Upon the substitution of this ansatz into equation (\ref{eq:eqS8}), we obtain the following system of the ordinary differential equations for functions $\mu(t)$ and $\nu(t)$:
\begin{equation}\label{eq:eqS14}
\left\{ 
\begin{array}{lr}
\dot{\mu} = -2 k_B T/\zeta \mu^2 + 2\kappa/\zeta \mu \\
\dot{\nu} = k_B T/\zeta \mu + k \langle X \rangle^2 - \kappa/\zeta
\end{array}
\right. 
\end{equation}
The first equation (\ref{eq:eqS14}) can be readily integrated to obtain the solution for $\mu(t)$. 
The obtained solution can be substituted into the second equation (\ref{eq:eqS14}), which can then be solved for $\nu(t)$. 
We obtain:
\begin{equation}\label{eq:eqS15}
\left\{ 
\begin{array}{lr}
\mu(t) = \frac{1}{C_1 e^{-2 t/\eta} + k_B T / \kappa} \\
\nu(t) = \frac{1}{2} \log \left[ k_B T + C_1 \kappa e^{-2 t/\eta} \right] + k \langle X \rangle^2 + C_2
\end{array}
\right. 
\end{equation}
When substitute into equation (\ref{eq:eqS13}), arrive at the expression:
\begin{equation}\label{eq:eqS16}
P(X, X_0; t) = \left[ \frac{2 \pi k_B T}{\kappa}  (1 - e^{-2t/\eta})\right]^{-1/2} \exp \left[ -\frac{\kappa (X - \langle X \rangle)^2}{2 k_B T (1 - e^{-2t/\eta})} \right] k \langle X \rangle^2 \exp \left[-k \langle X \rangle^2 t  \right] 
\end{equation}
which is equation (\ref{eq:eqS9}) above.

\newpage
\begin{table}[ht]
\caption{\label{tab:tabS1} Statistical analysis of experimentally determined bond lifetimes for different interacting protein-coated surfaces (specific versus non-specific protein-protein interactions) collected at a constant pulling force $f =$ 30 pN and contact duration $T =$ 0.5 s. Shown are the percentages of the bond lifetimes for the following time ranges: $\tau <$ 0.03 s, 0.04 s $< \tau <$ 0.5 s and $\tau >$ 0.5 s.}
\centering
\begin{tabular}{|c|c|c|c|c|}
\hline
\multirow{2}{*}{Interacting proteins} & \multicolumn{3}{ c| }{Bond lifetime ranges} & \multirow{2}{*}{No. of touching events}  \\ \cline{2-4}
 & $<$0.03 s & 0.04 - 0.5 s & $>$0.5 s &   \\ \hline
Fibrinogen/BSA & 96.9\% & 2.5\% & 0.6\%	& 4,136  \\ \hline
Fibrin/BSA & 98.1\% & 1.1\% & 0.8\% & 3,170  \\ \hline
Fibrinogen/Fibrinogen & 96.7\% & 2.0\% & 1.4\% & 3,481  \\ \hline
\cellcolor[gray]{0.9}Fibrin/Fibrinogen & 93.0\% & 0.6\% & \cellcolor[gray]{0.9}6.5\% & 4,125  \\ \hline
Fibrin/Fibrinogen in the presence of 2 mM GPRPam & 92.6\% & 6.1\% & 1.3\% & 4,023  \\ \hline
\end{tabular}
\end{table}

\begin{table}[ht]
\caption{\label{tab:tabS2} 
The model parameters for the A:a knob-hole bond dissociation kinetics: the binding interface stiffness $\kappa$, the initial (maximal) and minimal interface size (diameter) $x_0$ and $x_1$, the force-free rate of disruption of binding (receptor-ligand) residue-residue contacts per unit area $k_0$, transition distance for dissociation $\sigma_y$, and transition distance for conformational transition state $\sigma_x$ and energy difference between two states $\varepsilon_0$ (for discrete two-state version).
The values were obtained by fitting the theoretical profiles of the average bond lifetime $\langle \tau(f) \rangle$ as a function of applied constant force $f$ (curves) (equation (\ref{eq:eq7})) to the experimental bond lifetime data (points; see Fig.~ \ref{fig:fig5}). 
We used both continuous version (equation (\ref{eq:eqS2})) and discrete version (equation (\ref{eq:eqS3})) of the fluctuating bottleneck model of the receptor-ligand binding interface with a linear `sink' $K \sim X$ ($\alpha =$ 1; see equation (\ref{eq:eq2})).}
\centering
\begin{tabular}{|c|c|c|c|c|c|c|c|}
\hline
Model & $\kappa$, pN/nm & $x_0$, nm & $x_1$, nm & $k_0$, nm$^2$s$^{-1}$ & $\sigma_y$, nm & $\sigma_x$, nm & $\varepsilon_0$, kcal/mol \\
\hline
Continuous & 10.0 & 3.7 & 0.7 & 0.25 & 0.175 & - & - \\
\hline
Discrete & 100.0 & 5.0 & 0.65 & 0.095 & 0.195 & 17.0 & 1.242  \\
\hline
\end{tabular}
\end{table}

\newpage
\begin{figure}[ht]
\centering
\includegraphics[width=0.7\linewidth]{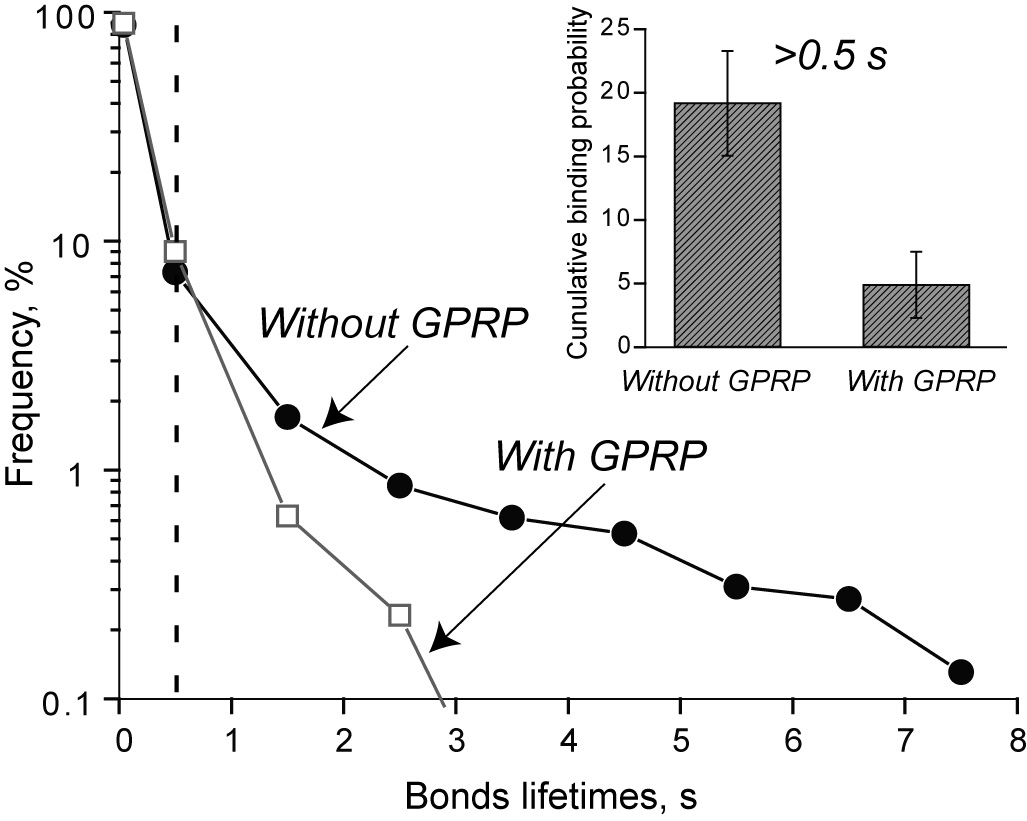}
\caption{
The log-linear plot of the distributions of bond lifetimes $P(t)$ collected at a constant pulling force $f =$ 50 pN for the fibrin-fibrinogen complexes in the absence and presence of 1 mM GPRP, a competitive inhibitor of the A:a knob:hole interactions. 
The vertical dashed line shows the bond lifetime threshold at 0.5 s, above which the A:a knob:hole interactions were found to be sensitive to the presence of GPRP peptide. 
Therefore, these data were considered to reflect the fibrin-fibrinogen specific interactions. 
The \textit{inset} compares the probability of interactions lasting longer than 0.5 s in the absence and presence of GPRP peptide.
}
\label{fig:figS1}
\end{figure}

\newpage
\begin{figure}[ht]
\centering
\includegraphics[width=0.7\linewidth]{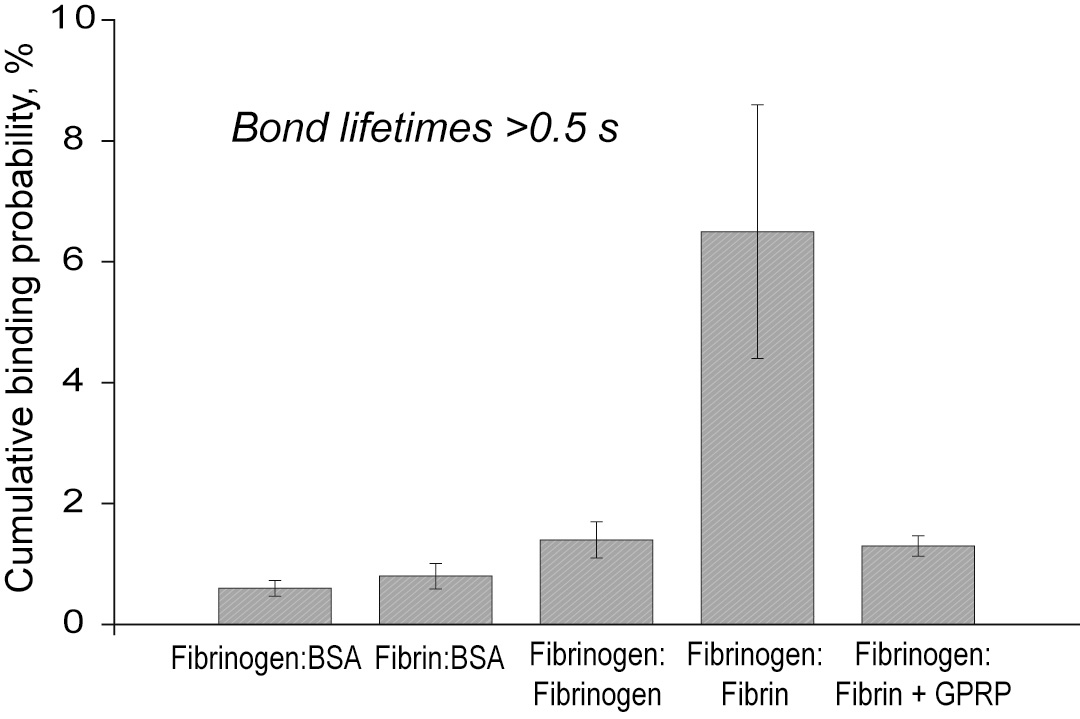}
\caption{
The binding probability for different interacting protein pairs showing that fibrin-fibrinogen complexes with lifetimes $\tau >$ 0.5 s collected at $f =$ 50 pN formed much more frequently and that the formation of these complexes is prevented by GPRP.
}
\label{fig:figS2}
\end{figure}

\newpage
\begin{figure}[ht]
\centering
\includegraphics[width=0.7\linewidth]{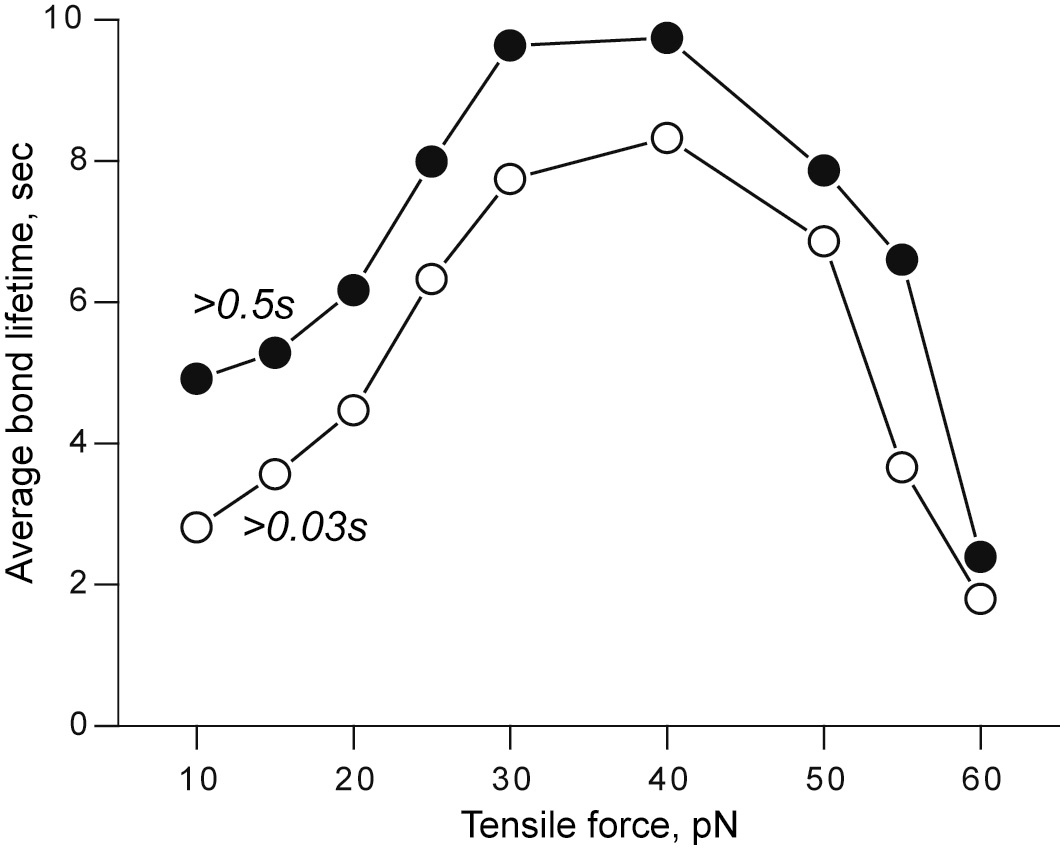}
\caption{
The average bond lifetimes $\langle \tau \rangle$ vs. applied constant force $f$ for specific fibrin-fibrinogen interactions calculated with the cutoff set at 0.03 s and 0.5 s.
}
\label{fig:figS3}
\end{figure}

\newpage
\begin{figure}[ht]
\centering
\includegraphics[width=0.7\linewidth]{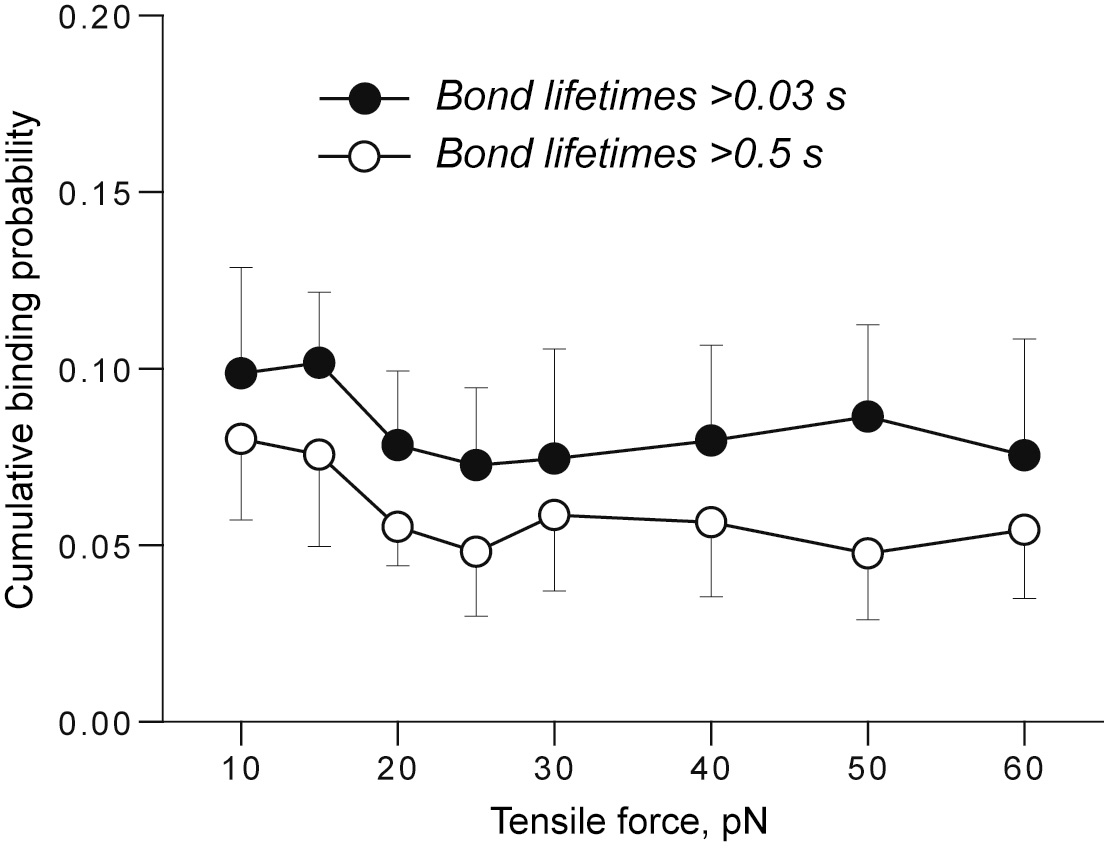}
\caption{
The binding probability as a function of the constant force f for fibrin-fibrinogen interactions calculated using the average lifetimes $\langle \tau \rangle >$ 0.03 s and $\langle \tau \rangle >$ 0.5 s.
}
\label{fig:figS4}
\end{figure}

\newpage
\begin{figure}[ht]
\centering
\includegraphics[width=0.7\linewidth]{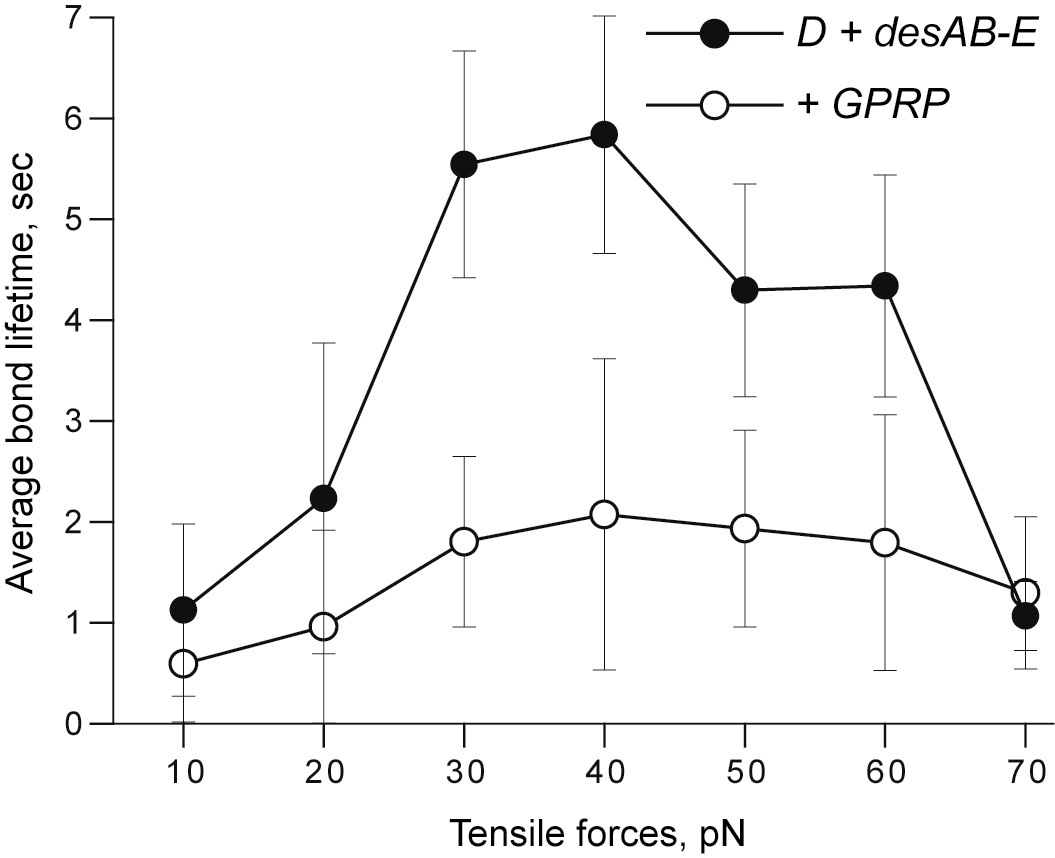}
\caption{
The average bond lifetime $\langle \tau \rangle >$ 0.5 s as a function of the constant tensile forces $f$ for interactions of fibrinogen fragment D (with holes `a' and `b') with fragment desAB-E (bearing knobs `A' and `B') in the absence and presence of 1 mM GPRP.
}
\label{fig:figS5}
\end{figure}

\newpage
\begin{figure}[ht]
\centering
\includegraphics[width=0.7\linewidth]{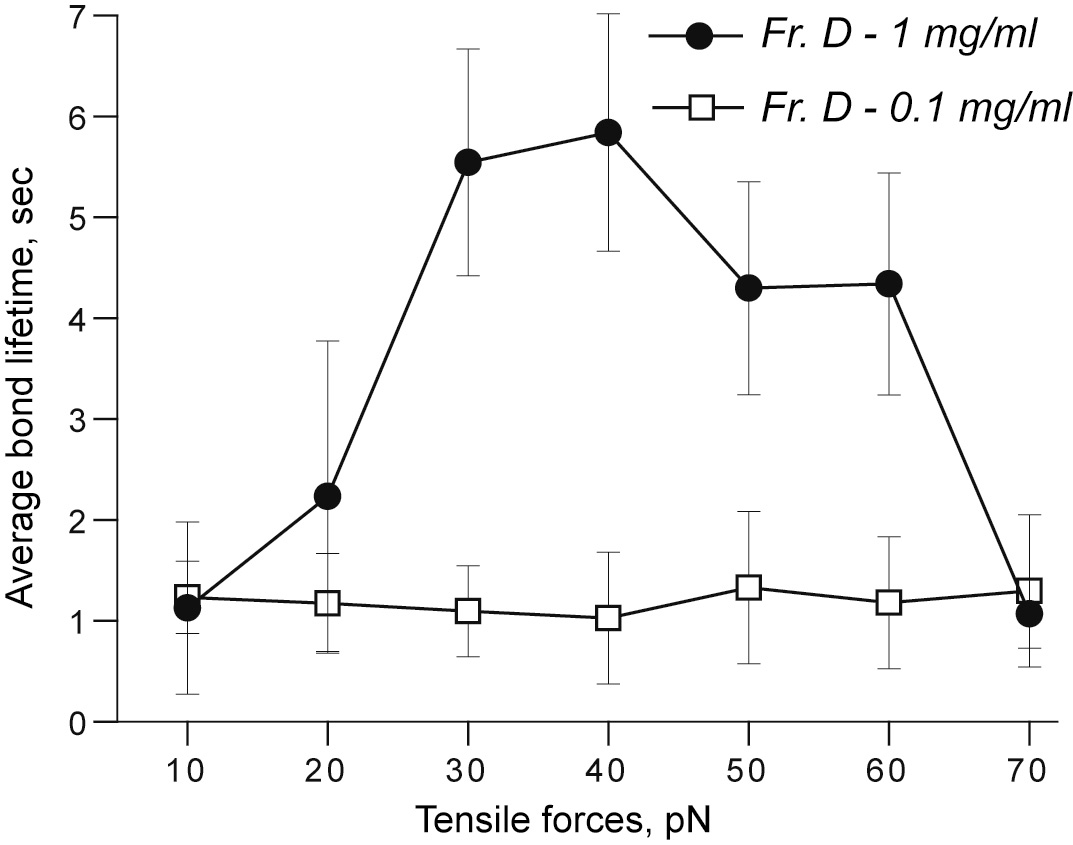}
\caption{
The average bond lifetime $\langle \tau \rangle >$ 0.5 s as a function of the constant tensile forces $f$ for interactions of fibrinogen fragment D (with holes `a' and `b') with fragment desAB-E (bearing knobs `A' and `B') obtained using 10-fold different concentrations of fragment D (1 mg/ml vs. 0.1 mg/ml). A 10-fold difference in fragment D concentration translates to different surface densities achieved during fragment D immobilization.
}
\label{fig:figS6}
\end{figure}

\newpage
\begin{figure}[ht]
\centering
\includegraphics[width=0.9\linewidth]{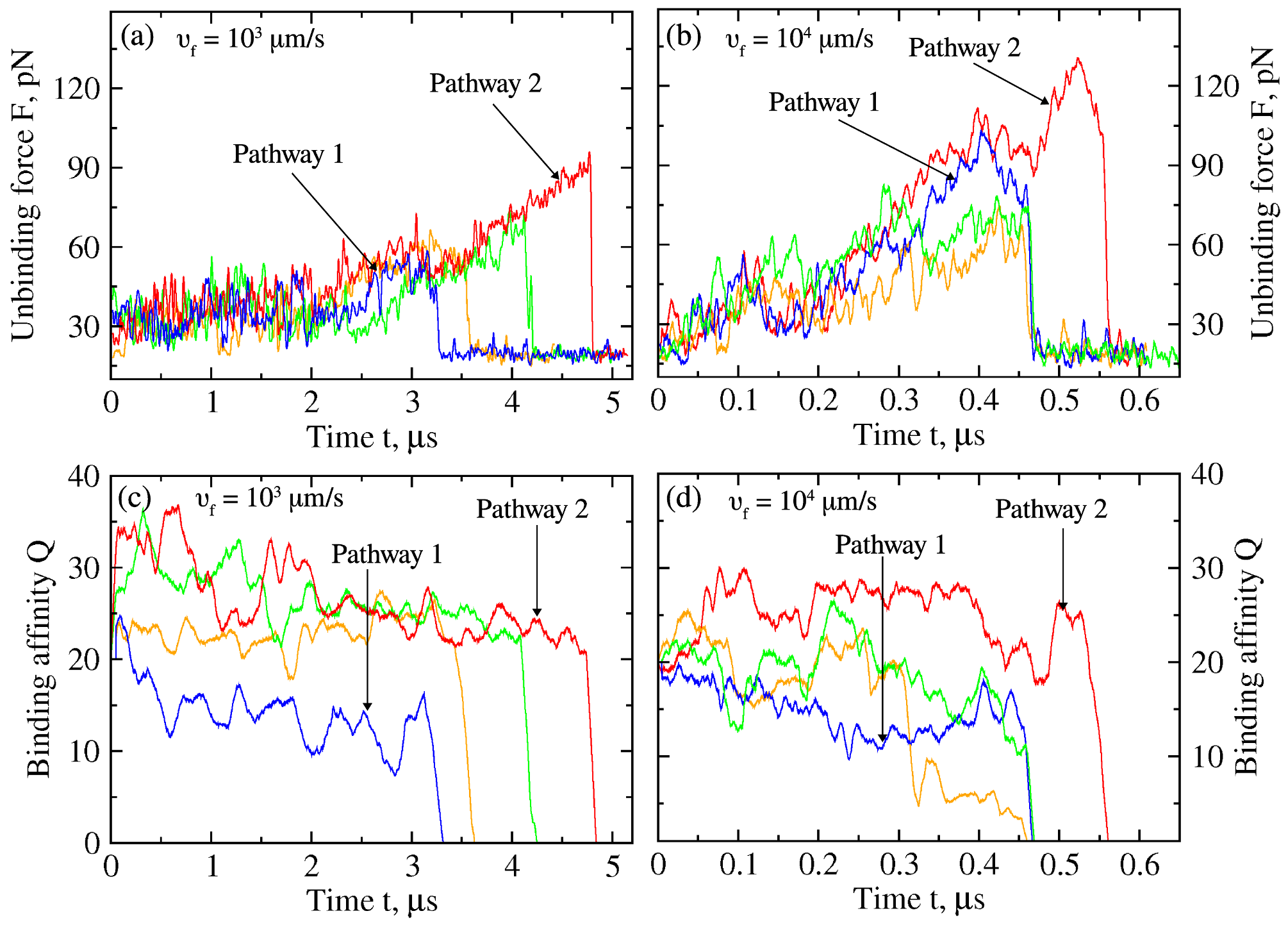}
\caption{
Dynamics of the A:a knob-hole interaction probed by varying mechanical tension. Shown are four representative profiles of the molecular (unbinding) force $F$ (panels a and b) and the number of residue-residue binding contacts reinforcing the bound state $Q$ (panels c and d) as functions of time $t$ from two sets of the force-ramp simulations with pulling speed $\nu_f = 10^3$ $\mu$m/s (panels a and c) and $\nu_f = 10^4$ $\mu$m/s (panels b and d). 
The text notations designate the most representative trajectories displaying the complex dissociation from the low-affinity bound state (pathway 1; blue curves) and high-affinity bound state (pathway 2; red curves).
}
\label{fig:figS7}
\end{figure}

\newpage
\begin{figure}[ht]
\centering
\includegraphics[width=0.9\linewidth]{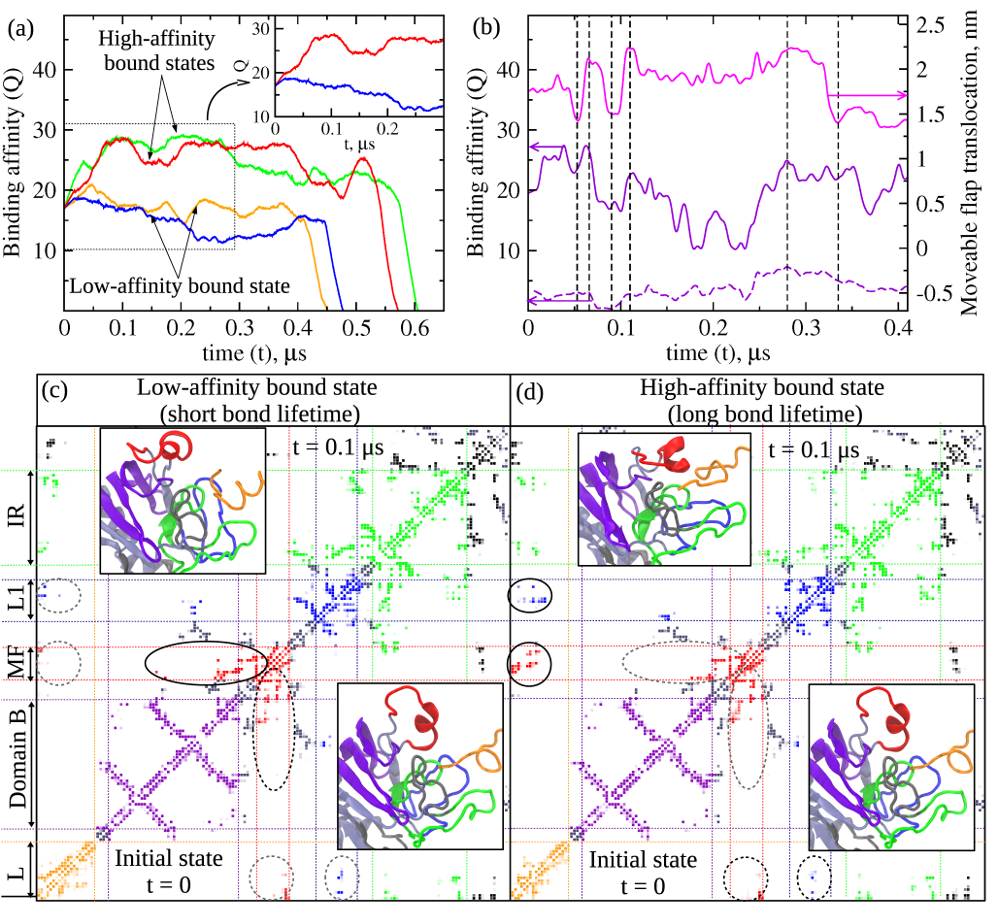}
\caption{
\textbf{Binding affinity of the A:a knob-hole complex.}
Panel a: Time evolution of the total number of binding contacts $Q$ from four representative force-ramp trajectories \textit{in silico} (the \textit{inset} magnifies initial portion of the curves, showing an increase or decrease in $Q$). 
Panel b: Time evolution of $Q$ (purple solid), number of binding contacts between moveable flap and knob `A' $Q_{MT}$ (purple dashed), and movable flap displacement relative to the B-domain $D_{MT}$ (magenta) from force-clamp simulations with $f =$ 30 pN (vertical dashed lines emphasize correlated alterations in $Q$ and $D_{MT}$). 
Panel c-d: Maps of residue-residue contacts stabilizing the complex interface (residues $\gamma$Ser240--$\gamma$Lys380) in the native state (bottom triangle) and intermediate state before dissociation (top triangle) for two simulations (blue and red curves in panel c) showing dissociation from the low-affinity bound state (panel c) and high-affinity state (panel d). 
A colored pixel corresponds to a contact, formed by amino acids in the $i$-th raw and $j$-th column, if the centers-of-mass of their side chains are in 5.5 \AA~proximity. 
The color opacity is proportional to the contacts persistence. 
Different colors denote contacts formed by residues in knob `A' (orange) and residues in one of the binding determinants in $\gamma$-nodule: loop I (blue), interior region (green), and movable flap (red). 
The $\beta$-sheets stack of B-domain is shown in purple. 
The solid ovals circle the areas of new contacts formed between the movable flap and $\beta$-sheet stack in B-domain (left) and between the movable flap and knob `A' (right), which the native state lacks (dashed oval).
}
\label{fig:figS8}
\end{figure}

\newpage
\begin{figure}[ht]
\centering
\includegraphics[width=0.9\linewidth]{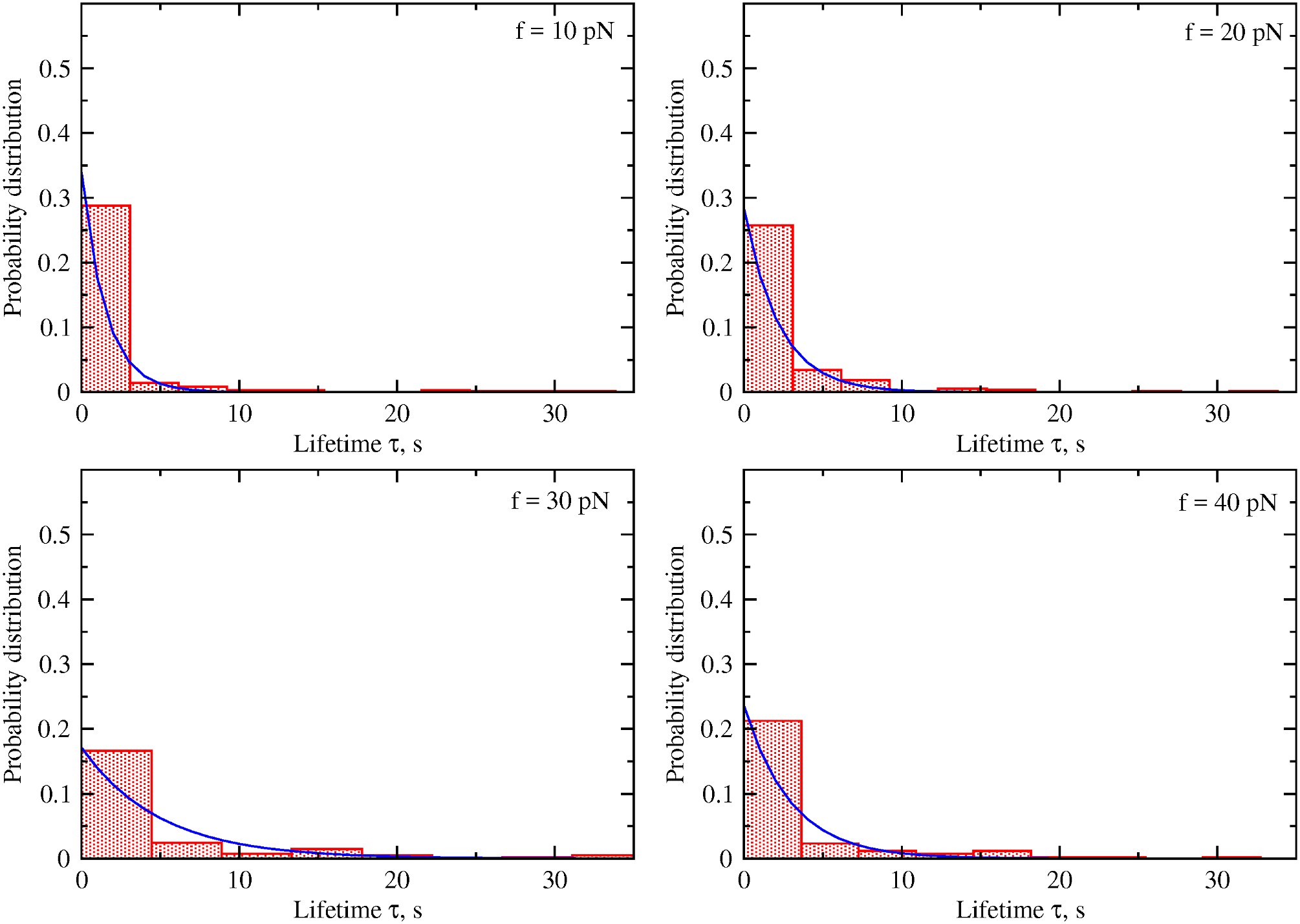}
\caption{
Probability distributions of bond lifetimes obtained for different values of tensile force $f =$ 10, 20, 30, 40 pN. Comparison of experimental (red histograms) vs. modeling (blue curves) results (equation \ref{eq:eq6}).
}
\label{fig:figS9}
\end{figure}

%% file: Litvinov_et_al_CatchSlip.bbl
\begin{thebibliography}{10}
\expandafter\ifx\csname url\endcsname\relax
  \def\url#1{\texttt{#1}}\fi
\expandafter\ifx\csname urlprefix\endcsname\relax\def\urlprefix{URL }\fi
\expandafter\ifx\csname doiprefix\endcsname\relax\def\doiprefix{DOI }\fi
\providecommand{\bibinfo}[2]{#2}
\providecommand{\eprint}[2][]{\url{#2}}

\bibitem{Litvinov&WeiselSTH16}
\bibinfo{author}{Litvinov, R.~I.} \& \bibinfo{author}{Weisel, J.~W.}
\newblock \bibinfo{journal}{\bibinfo{title}{What is the biological and clinical
  relevance of fibrin?}}
\newblock {\emph{\JournalTitle{Semin. Thromb. Hemost.}}}
  \textbf{\bibinfo{volume}{42}}, \bibinfo{pages}{333--343}
  (\bibinfo{year}{2016}).

\bibitem{Weisel&LitvinovSB17}
\bibinfo{author}{Weisel, J.~W.} \& \bibinfo{author}{Litvinov, R.~I.}
\newblock \bibinfo{journal}{\bibinfo{title}{Fibrin formation, structure and
  properties}}.
\newblock {\emph{\JournalTitle{Subcell Biochem.}}}
  \textbf{\bibinfo{volume}{82}}, \bibinfo{pages}{405--456}
  (\bibinfo{year}{2017}).

\bibitem{StandevenBR05}
\bibinfo{author}{Standeven, K.~F.}, \bibinfo{author}{Arieons, R. A.~S.} \&
  \bibinfo{author}{Grant, P.~J.}
\newblock \bibinfo{journal}{\bibinfo{title}{The molecular physiology and
  pathology of fibrin structure/function}}.
\newblock {\emph{\JournalTitle{Blood Rev.}}} \textbf{\bibinfo{volume}{19}},
  \bibinfo{pages}{275--288} (\bibinfo{year}{2005}).

\bibitem{LaudanoPNAS78}
\bibinfo{author}{Laudano, A.~P.} \& \bibinfo{author}{Doolittle, R.~F.}
\newblock \bibinfo{journal}{\bibinfo{title}{Synthetic peptide derivatives that
  bind to fibrinogen and prevent the polymerization of fibrin monomers}}.
\newblock {\emph{\JournalTitle{Proc. Natl. Acad. Sci. USA}}}
  \textbf{\bibinfo{volume}{75}}, \bibinfo{pages}{3085--3089}
  (\bibinfo{year}{1978}).

\bibitem{LaudanoBiochemistry80}
\bibinfo{author}{Laudano, A.~P.} \& \bibinfo{author}{Doolittle, R.~F.}
\newblock \bibinfo{journal}{\bibinfo{title}{Studies on synthetic peptides that
  bind to fibrinogen and prevent fibrin polymerization. {Structural}
  requirements, number of binding sites, and species differences}}.
\newblock {\emph{\JournalTitle{Biochemistry}}} \textbf{\bibinfo{volume}{19}},
  \bibinfo{pages}{1013--1019} (\bibinfo{year}{1980}).

\bibitem{SpraggonNature97}
\bibinfo{author}{Spraggon, G.}, \bibinfo{author}{Everse, S.~J.} \&
  \bibinfo{author}{Doolittle, R.~F.}
\newblock \bibinfo{journal}{\bibinfo{title}{Crystal structures of fragment {D}
  from human fibrinogen and its crosslinked counterpart from fibrin}}.
\newblock {\emph{\JournalTitle{Nature}}} \textbf{\bibinfo{volume}{389}},
  \bibinfo{pages}{455--462} (\bibinfo{year}{1997}).

\bibitem{LitvinovBlood05}
\bibinfo{author}{Litvinov, R.~I.}, \bibinfo{author}{Gorkun, O.~V.},
  \bibinfo{author}{Owen, S.~F.}, \bibinfo{author}{Shuman, H.} \&
  \bibinfo{author}{Weisel, J.~W.}
\newblock \bibinfo{journal}{\bibinfo{title}{Polymerization of fibrin:
  specificity, strength, and stability of knob-hole interactions studied at the
  single-molecule level}}.
\newblock {\emph{\JournalTitle{Blood}}} \textbf{\bibinfo{volume}{106}},
  \bibinfo{pages}{2944--2951} (\bibinfo{year}{2005}).

\bibitem{LitvinovJTH13}
\bibinfo{author}{Litvinov, R.~I.} \& \bibinfo{author}{Weisel, J.~W.}
\newblock \bibinfo{journal}{\bibinfo{title}{Shear strengthens fibrin: the
  knob–hole interactions display `catch-slip' kinetics}}.
\newblock {\emph{\JournalTitle{J. Thromb. Haemost.}}}
  \textbf{\bibinfo{volume}{11}}, \bibinfo{pages}{1933--1935}
  (\bibinfo{year}{2013}).

\bibitem{HertigCurrBiol12}
\bibinfo{author}{Hertig, S.} \& \bibinfo{author}{Vogel, V.}
\newblock \bibinfo{journal}{\bibinfo{title}{Catch bonds}}.
\newblock {\emph{\JournalTitle{Curr. Biol.}}} \textbf{\bibinfo{volume}{22}},
  \bibinfo{pages}{R823} (\bibinfo{year}{2012}).

\bibitem{MarshallNat03}
\bibinfo{author}{Marshall, B.~T.} \emph{et~al.}
\newblock \bibinfo{journal}{\bibinfo{title}{Direct observation of catch bonds
  involving cell-adhesion molecules}}.
\newblock {\emph{\JournalTitle{Nature}}} \textbf{\bibinfo{volume}{423}},
  \bibinfo{pages}{190--193} (\bibinfo{year}{2003}).

\bibitem{RakshitaPNAS12}
\bibinfo{author}{Rakshita, S.}, \bibinfo{author}{Zhang, Y.},
  \bibinfo{author}{Manibog, K.}, \bibinfo{author}{Shafraza, O.} \&
  \bibinfo{author}{Sivasankar, S.}
\newblock \bibinfo{journal}{\bibinfo{title}{Ideal, catch, and slip bonds in
  cadherin adhesion}}.
\newblock {\emph{\JournalTitle{Proc. Natl. Acad. Sci. USA}}}
  \textbf{\bibinfo{volume}{106}}, \bibinfo{pages}{18815--18820}
  (\bibinfo{year}{2012}).

\bibitem{KongJCB09}
\bibinfo{author}{Kong, F.}, \bibinfo{author}{Garcia, A.~J.},
  \bibinfo{author}{Mould, A.~P.}, \bibinfo{author}{Humphries, M.~J.} \&
  \bibinfo{author}{Zhu, C.}
\newblock \bibinfo{journal}{\bibinfo{title}{Demonstration of catch bonds
  between an integrin and its ligand}}.
\newblock {\emph{\JournalTitle{J. Cell Biol.}}} \textbf{\bibinfo{volume}{185}},
  \bibinfo{pages}{1275--1284} (\bibinfo{year}{2009}).

\bibitem{ThomasCell02}
\bibinfo{author}{Thomas, W.~E.}, \bibinfo{author}{Trintchina, E.},
  \bibinfo{author}{Forero, M.}, \bibinfo{author}{Vogel, V.} \&
  \bibinfo{author}{Sokurenko, E.~V.}
\newblock \bibinfo{journal}{\bibinfo{title}{Bacterial adhesion to target cells
  enhanced by shear force}}.
\newblock {\emph{\JournalTitle{Cell}}} \textbf{\bibinfo{volume}{109}},
  \bibinfo{pages}{913--923} (\bibinfo{year}{2002}).

\bibitem{SauerNC16}
\bibinfo{author}{Sauer, M.~M.} \emph{et~al.}
\newblock \bibinfo{journal}{\bibinfo{title}{Catch-bond mechanism of the
  bacterial adhesin {FimH}}}.
\newblock {\emph{\JournalTitle{Nature Comm.}}} \textbf{\bibinfo{volume}{7}},
  \bibinfo{pages}{10738} (\bibinfo{year}{2016}).

\bibitem{YagoJCI08}
\bibinfo{author}{Yago, T.} \emph{et~al.}
\newblock \bibinfo{journal}{\bibinfo{title}{Platelet glycoprotein {Ib$\alpha$}
  forms catch bonds with human {WT} {vWF} but not with type {2B} von
  {Willebrand} disease {vWF}}}.
\newblock {\emph{\JournalTitle{J. Clin. Invest.}}}
  \textbf{\bibinfo{volume}{118}}, \bibinfo{pages}{3195--3207}
  (\bibinfo{year}{2006}).

\bibitem{FeghhiBJ16}
\bibinfo{author}{Feghhi, S.} \emph{et~al.}
\newblock \bibinfo{journal}{\bibinfo{title}{Glycoprotein {Ib-IX-V} complex
  transmits cytoskeletal forces that enhance platelet adhesion}}.
\newblock {\emph{\JournalTitle{Biophys. J.}}} \textbf{\bibinfo{volume}{111}},
  \bibinfo{pages}{601--608} (\bibinfo{year}{2016}).

\bibitem{GuoPNAS06}
\bibinfo{author}{Guo, B.} \& \bibinfo{author}{Guilford, W.~H.}
\newblock \bibinfo{journal}{\bibinfo{title}{Mechanics of actomyosin bonds in
  different nucleotide states are tuned to muscle contraction}}.
\newblock {\emph{\JournalTitle{Proc. Natl. Acad. Sci. USA}}}
  \textbf{\bibinfo{volume}{103}}, \bibinfo{pages}{9844--9849}
  (\bibinfo{year}{2006}).

\bibitem{AkiyoshiNature10}
\bibinfo{author}{Akiyoshi, B.} \emph{et~al.}
\newblock \bibinfo{journal}{\bibinfo{title}{Tension directly stabilizes
  reconstituted kinetochore-microtubule attachments}}.
\newblock {\emph{\JournalTitle{Nature}}} \textbf{\bibinfo{volume}{468}},
  \bibinfo{pages}{576--579} (\bibinfo{year}{2010}).

\bibitem{RaiCell13}
\bibinfo{author}{Rai, A.~K.}, \bibinfo{author}{Rai, A.},
  \bibinfo{author}{Ramaiya, A.~J.}, \bibinfo{author}{Jha, R.} \&
  \bibinfo{author}{Mallik, R.}
\newblock \bibinfo{journal}{\bibinfo{title}{Molecular adaptations allow dynein
  to generate large collective forces inside cells}}.
\newblock {\emph{\JournalTitle{Cell}}} \textbf{\bibinfo{volume}{152}},
  \bibinfo{pages}{172--182} (\bibinfo{year}{2013}).

\bibitem{NairPRE}
\bibinfo{author}{Nair, A.}, \bibinfo{author}{Chandel, S.},
  \bibinfo{author}{Mitra, M.~K.}, \bibinfo{author}{Muhuri, S.} \&
  \bibinfo{author}{Chaudhuri, A.}
\newblock \bibinfo{journal}{\bibinfo{title}{Effect of catch bonding on
  transport of cellular cargo by dynein motors}}.
\newblock {\emph{\JournalTitle{Phys. Rev. E}}} \textbf{\bibinfo{volume}{94}},
  \bibinfo{pages}{032403} (\bibinfo{year}{2016}).

\bibitem{ChenJBC10}
\bibinfo{author}{Chen, W.}, \bibinfo{author}{Lou, J.} \& \bibinfo{author}{Zhu,
  C.}
\newblock \bibinfo{journal}{\bibinfo{title}{Forcing switch from short- to
  intermediate- and long-lived states of the $\alpha${A} domain generates
  {LFA-1/ICAM-1} catch bonds}}.
\newblock {\emph{\JournalTitle{J. Biol. Chem.}}}
  \textbf{\bibinfo{volume}{285}}, \bibinfo{pages}{35967--35978}
  (\bibinfo{year}{2010}).

\bibitem{McEverARCDB10}
\bibinfo{author}{McEver, R.~P.} \& \bibinfo{author}{Zhu, C.}
\newblock \bibinfo{journal}{\bibinfo{title}{Rolling cell adhesion}}.
\newblock {\emph{\JournalTitle{Annu. Rev. Cell Dev. Biol.}}}
  \textbf{\bibinfo{volume}{26}}, \bibinfo{pages}{363--396}
  (\bibinfo{year}{2010}).

\bibitem{WaldronPNAS08}
\bibinfo{author}{Waldron, T.~T.} \& \bibinfo{author}{Springer, T.~A.}
\newblock \bibinfo{journal}{\bibinfo{title}{Transmission of allostery through
  the lectin domain in selectin-mediated cell adhesion}}.
\newblock {\emph{\JournalTitle{Proc. Natl. Acad. Sci. USA}}}
  \textbf{\bibinfo{volume}{106}}, \bibinfo{pages}{85--90}
  (\bibinfo{year}{2008}).

\bibitem{SarangapaniJBC11}
\bibinfo{author}{Sarangapani, K.~K.} \emph{et~al.}
\newblock \bibinfo{journal}{\bibinfo{title}{Regulation of catch bonds by rate
  of force application}}.
\newblock {\emph{\JournalTitle{J. Biol. Chem.}}}
  \textbf{\bibinfo{volume}{286}}, \bibinfo{pages}{32749--32761}
  (\bibinfo{year}{2011}).

\bibitem{HelmsFEBSLet16}
\bibinfo{author}{Helms, G.}, \bibinfo{author}{Dasanna, A.~K.},
  \bibinfo{author}{Schwarz, U.~S.} \& \bibinfo{author}{Lanzer, M.}
\newblock \bibinfo{journal}{\bibinfo{title}{Modeling cytoadhesion of plasmodium
  falciparum-infected erythrocytes and leukocytes-common principles and
  distinctive features}}.
\newblock {\emph{\JournalTitle{FEBS Lett.}}} \textbf{\bibinfo{volume}{590}},
  \bibinfo{pages}{1955--1971} (\bibinfo{year}{2016}).

\bibitem{ManibogNC14}
\bibinfo{author}{Manibog, K.}, \bibinfo{author}{Li, H.},
  \bibinfo{author}{Rakshit, S.} \& \bibinfo{author}{Sivasankar, S.}
\newblock \bibinfo{journal}{\bibinfo{title}{Resolving the molecular mechanism
  of cadherin catch bond formation}}.
\newblock {\emph{\JournalTitle{Nature Comm.}}} \textbf{\bibinfo{volume}{5}},
  \bibinfo{pages}{3941} (\bibinfo{year}{2014}).

\bibitem{GunnersonJPCB09}
\bibinfo{author}{Gunnerson, K.~N.}, \bibinfo{author}{Pereverzev, Y.~V.} \&
  \bibinfo{author}{Prezhdo, O.~V.}
\newblock \bibinfo{journal}{\bibinfo{title}{Atomistic simulation combined with
  analytic theory to study the response of the {P}-selectin/{PSGL-1} complex to
  an external force}}.
\newblock {\emph{\JournalTitle{J. Phys. Chem. B}}}
  \textbf{\bibinfo{volume}{113}}, \bibinfo{pages}{2090--2100}
  (\bibinfo{year}{2009}).

\bibitem{ChakrabartiJSB17}
\bibinfo{author}{Chakrabarti, S.}, \bibinfo{author}{Hinczewski, M.} \&
  \bibinfo{author}{Thirumalai, D.}
\newblock \bibinfo{journal}{\bibinfo{title}{Phenomenological and microscopic
  theories for catch bonds}}.
\newblock {\emph{\JournalTitle{J. Struct. Biol.}}}
  \textbf{\bibinfo{volume}{197}}, \bibinfo{pages}{50--56}
  (\bibinfo{year}{2017}).

\bibitem{Vernerey&AkalpPRE16}
\bibinfo{author}{Vernerey, F.~J.} \& \bibinfo{author}{Akalp, U.}
\newblock \bibinfo{journal}{\bibinfo{title}{Role of catch bonds in actomyosin
  mechanics and cell mechanosensitivity}}.
\newblock {\emph{\JournalTitle{Phys. Rev. E}}} \textbf{\bibinfo{volume}{94}},
  \bibinfo{pages}{012403} (\bibinfo{year}{2016}).

\bibitem{Bullerjahn&KroyPRE16}
\bibinfo{author}{Bullerjahn, J.~T.} \& \bibinfo{author}{Kroy, K.}
\newblock \bibinfo{journal}{\bibinfo{title}{Analytical catch-slip bond model
  for arbitrary forces and loading rates}}.
\newblock {\emph{\JournalTitle{Phys. Rev. E}}} \textbf{\bibinfo{volume}{93}},
  \bibinfo{pages}{012404} (\bibinfo{year}{2016}).

\bibitem{BarsegovPNAS05}
\bibinfo{author}{Barsegov, V.} \& \bibinfo{author}{Thirumalai, D.}
\newblock \bibinfo{journal}{\bibinfo{title}{Dynamics of unbinding of cell
  adhesion molecules: {Transition} from catch to slip bonds}}.
\newblock {\emph{\JournalTitle{Proc. Natl. Acad. Sci. U. S. A.}}}
  \textbf{\bibinfo{volume}{102}}, \bibinfo{pages}{1835--1839}
  (\bibinfo{year}{2005}).

\bibitem{BarsegovJPCB06}
\bibinfo{author}{Barsegov, V.} \& \bibinfo{author}{Thirumalai, D.}
\newblock \bibinfo{journal}{\bibinfo{title}{Dynamic competition between catch
  and slip bonds in selectins bound to ligands}}.
\newblock {\emph{\JournalTitle{J. Phys. Chem. B}}}
  \textbf{\bibinfo{volume}{110}}, \bibinfo{pages}{26403--26412}
  (\bibinfo{year}{2006}).

\bibitem{PereverzevBJ05}
\bibinfo{author}{Pereverzev, Y.~V.}, \bibinfo{author}{Prezhdo, O.~V.},
  \bibinfo{author}{Forero, M.}, \bibinfo{author}{Sokurenko, E.~V.} \&
  \bibinfo{author}{Thomas, W.~E.}
\newblock \bibinfo{journal}{\bibinfo{title}{The two-pathway model for the
  catch-slip transition in biological adhesion}}.
\newblock {\emph{\JournalTitle{Biophys. J.}}} \textbf{\bibinfo{volume}{89}},
  \bibinfo{pages}{91446--91454} (\bibinfo{year}{2005}).

\bibitem{LouBJ07}
\bibinfo{author}{Lou, J.} \& \bibinfo{author}{Zhu, C.}
\newblock \bibinfo{journal}{\bibinfo{title}{A structure-based sliding-rebinding
  mechanism for catch bonds}}.
\newblock {\emph{\JournalTitle{Biophys. J.}}} \textbf{\bibinfo{volume}{92}},
  \bibinfo{pages}{1471--1485} (\bibinfo{year}{2007}).

\bibitem{ChakrabartiPNAS14}
\bibinfo{author}{Chakrabarti, S.}, \bibinfo{author}{Hinczewski, M.} \&
  \bibinfo{author}{Thirumalai, D.}
\newblock \bibinfo{journal}{\bibinfo{title}{Plasticity of hydrogen bond
  networks regulates mechanochemistry of cell adhesion complexes}}.
\newblock {\emph{\JournalTitle{Proc. Natl. Acad. Sci. USA}}}
  \textbf{\bibinfo{volume}{111}}, \bibinfo{pages}{9048--9053}
  (\bibinfo{year}{2014}).

\bibitem{LiuPRE06}
\bibinfo{author}{Liu, F.} \& \bibinfo{author}{{Ou-Yang}, Z.}
\newblock \bibinfo{journal}{\bibinfo{title}{Force modulating dynamic disorder:
  {A} physical model of catch-slip bond transitions in receptor-ligand forced
  dissociation experiments}}.
\newblock {\emph{\JournalTitle{Phys. Rev. E}}} \textbf{\bibinfo{volume}{74}},
  \bibinfo{pages}{051904} (\bibinfo{year}{2006}).

\bibitem{WeiPRE08}
\bibinfo{author}{Wei, Y.}
\newblock \bibinfo{journal}{\bibinfo{title}{Entropic-elasticity-controlled
  dissociation and energetic-elasticity-controlled rupture induce catch-to-slip
  bonds in cell-adhesion molecules}}.
\newblock {\emph{\JournalTitle{Phys. Rev. E}}} \textbf{\bibinfo{volume}{77}},
  \bibinfo{pages}{031910} (\bibinfo{year}{2008}).

\bibitem{LitvinovBlood07}
\bibinfo{author}{Litvinov, R.~I.} \emph{et~al.}
\newblock \bibinfo{journal}{\bibinfo{title}{Polymerization of fibrin: direct
  observation and quantification of individual {B}:b knob-hole interactions}}.
\newblock {\emph{\JournalTitle{Blood}}} \textbf{\bibinfo{volume}{109}},
  \bibinfo{pages}{130--138} (\bibinfo{year}{2007}).

\bibitem{ZhmurovStructure16}
\bibinfo{author}{Zhmurov, A.} \emph{et~al.}
\newblock \bibinfo{journal}{\bibinfo{title}{Structural basis of interfacial
  flexibility in fibrin oligomers}}.
\newblock {\emph{\JournalTitle{Structure}}} \textbf{\bibinfo{volume}{24}},
  \bibinfo{pages}{1907--1917} (\bibinfo{year}{2016}).

\bibitem{ZwanzigJCP92}
\bibinfo{author}{Zwanzig, R.}
\newblock \bibinfo{journal}{\bibinfo{title}{Dynamical disorder: Passage through
  a fluctuating bottleneck}}.
\newblock {\emph{\JournalTitle{J. Chem. Phys.}}} \textbf{\bibinfo{volume}{97}},
  \bibinfo{pages}{3587} (\bibinfo{year}{1992}).

\bibitem{BarsegovJCP02}
\bibinfo{author}{Barsegov, V.}, \bibinfo{author}{Chernyak, V.} \&
  \bibinfo{author}{Mukamel, S.}
\newblock \bibinfo{journal}{\bibinfo{title}{Multitime correlation functions for
  single molecule kinetics with fluctuating bottlenecks}}.
\newblock {\emph{\JournalTitle{J. Chem. Phys.}}}
  \textbf{\bibinfo{volume}{116}}, \bibinfo{pages}{4240--4251}
  (\bibinfo{year}{2002}).

\bibitem{HyeonPRL14}
\bibinfo{author}{Hyeon, C.}, \bibinfo{author}{Hinczewski, M.} \&
  \bibinfo{author}{Thirumalai, D.}
\newblock \bibinfo{journal}{\bibinfo{title}{Evidence of disorder in biological
  molecules from single molecule pulling experiments}}.
\newblock {\emph{\JournalTitle{Phys. Rev. Lett.}}}
  \textbf{\bibinfo{volume}{112}}, \bibinfo{pages}{138101}
  (\bibinfo{year}{2014}).

\bibitem{Bicout&SzaboJCP98}
\bibinfo{author}{Bicout, D.~J.} \& \bibinfo{author}{Szabo, A.}
\newblock \bibinfo{journal}{\bibinfo{title}{Escape through a bottleneck
  undergoing non-markovian fluctuations}}.
\newblock {\emph{\JournalTitle{J. Chem. Phys.}}}
  \textbf{\bibinfo{volume}{108}}, \bibinfo{pages}{5491} (\bibinfo{year}{1998}).

\bibitem{Wang&WolynesCPL93}
\bibinfo{author}{Wang, J.} \& \bibinfo{author}{Wolynes, P.}
\newblock \bibinfo{journal}{\bibinfo{title}{Passage through fluctuating
  geometrical bottlenecks. {T}he general gaussian fluctuating case}}.
\newblock {\emph{\JournalTitle{Chem. Phys. Lett.}}}
  \textbf{\bibinfo{volume}{212}}, \bibinfo{pages}{427--433}
  (\bibinfo{year}{1993}).

\bibitem{BellScience78}
\bibinfo{author}{Bell, G.~L.}
\newblock \bibinfo{journal}{\bibinfo{title}{Models for the specific adhesion of
  cells to cells}}.
\newblock {\emph{\JournalTitle{Science}}} \textbf{\bibinfo{volume}{200}},
  \bibinfo{pages}{618--627} (\bibinfo{year}{1978}).

\bibitem{SavageCell96}
\bibinfo{author}{Savage, B.}, \bibinfo{author}{Saldivar, E.} \&
  \bibinfo{author}{Ruggeri, Z.~M.}
\newblock \bibinfo{journal}{\bibinfo{title}{Initiation of platelet adhesion by
  arrest onto fibrinogen or translocation on {von Willebrand} factor}}.
\newblock {\emph{\JournalTitle{Cell}}} \textbf{\bibinfo{volume}{84}},
  \bibinfo{pages}{289--297} (\bibinfo{year}{1996}).

\bibitem{PrezhdoACR09}
\bibinfo{author}{Prezhdo, O.~V.} \& \bibinfo{author}{Pereverzev, Y.~V.}
\newblock \bibinfo{journal}{\bibinfo{title}{Theoretical aspects of the
  biological catch bond}}.
\newblock {\emph{\JournalTitle{Acc. Chem. Res.}}}
  \textbf{\bibinfo{volume}{142}}, \bibinfo{pages}{693--703}
  (\bibinfo{year}{2009}).

\bibitem{ThomasCOSB09}
\bibinfo{author}{Thomas, W.~E.}
\newblock \bibinfo{journal}{\bibinfo{title}{Mechanochemistry of receptor-ligand
  bonds}}.
\newblock {\emph{\JournalTitle{Curr. Opin. Struct. Biol.}}}
  \textbf{\bibinfo{volume}{19}}, \bibinfo{pages}{50--55}
  (\bibinfo{year}{2009}).

\bibitem{ZhuBiorheol05}
\bibinfo{author}{Zhu, C.}, \bibinfo{author}{Lou, J.} \&
  \bibinfo{author}{McEver, R.~P.}
\newblock \bibinfo{journal}{\bibinfo{title}{Catch bonds: physical models,
  structural bases, biological function and rheological relevance}}.
\newblock {\emph{\JournalTitle{Biorheology}}} \textbf{\bibinfo{volume}{42}},
  \bibinfo{pages}{443--462} (\bibinfo{year}{2005}).

\bibitem{Chen&AlexanderKatzBJ11}
\bibinfo{author}{Chen, H.} \& \bibinfo{author}{{Alexander-Katz}, A.}
\newblock \bibinfo{journal}{\bibinfo{title}{Polymer-based catch-bonds}}.
\newblock {\emph{\JournalTitle{Biophys. J.}}} \textbf{\bibinfo{volume}{100}},
  \bibinfo{pages}{174--182} (\bibinfo{year}{2011}).

\bibitem{FalkovichPSSA10}
\bibinfo{author}{Falkovich, S.}, \bibinfo{author}{Neelov, I.} \&
  \bibinfo{author}{Darinskii, A.}
\newblock \bibinfo{journal}{\bibinfo{title}{Mechanism of shear deformation of a
  coiled myosin coil: Computer simulation}}.
\newblock {\emph{\JournalTitle{Polym. Sci. Ser. A+}}}
  \textbf{\bibinfo{volume}{62}}, \bibinfo{pages}{662} (\bibinfo{year}{2010}).

\bibitem{LitvinovPNAS02}
\bibinfo{author}{Litvinov, R.~I.}, \bibinfo{author}{Shuman, H.},
  \bibinfo{author}{Bennett, J.~S.} \& \bibinfo{author}{Weisel, J.~W.}
\newblock \bibinfo{journal}{\bibinfo{title}{Binding strength and activation
  state of single fibrinogen-integrin pairs on living cells}}.
\newblock {\emph{\JournalTitle{Proc. Natl. Acad. Sci. USA}}}
  \textbf{\bibinfo{volume}{99}}, \bibinfo{pages}{7426--7431}
  (\bibinfo{year}{2002}).

\bibitem{LitvinovBJ05}
\bibinfo{author}{Litvinov, R.~I.}, \bibinfo{author}{Bennett, J.~S.},
  \bibinfo{author}{Weisel, J.~W.} \& \bibinfo{author}{Shuman, H.}
\newblock \bibinfo{journal}{\bibinfo{title}{Multi-step fibrinogen binding to
  the integrin {$\alpha$IIb$\beta$3} detected using force spectroscopy}}.
\newblock {\emph{\JournalTitle{Biophys. J.}}} \textbf{\bibinfo{volume}{89}},
  \bibinfo{pages}{2824--2834} (\bibinfo{year}{2005}).

\bibitem{LitvinovBJ11}
\bibinfo{author}{Litvinov, R.~I.} \emph{et~al.}
\newblock \bibinfo{journal}{\bibinfo{title}{Dissociation of bimolecular
  {$\alpha$IIb$\beta$3}-fibrinogen complex under a constant tensile force}}.
\newblock {\emph{\JournalTitle{Biophys. J.}}} \textbf{\bibinfo{volume}{100}},
  \bibinfo{pages}{165--173} (\bibinfo{year}{2011}).

\bibitem{LitvinovJBC12}
\bibinfo{author}{Litvinov, R.~I.} \emph{et~al.}
\newblock \bibinfo{journal}{\bibinfo{title}{Resolving two-dimensional kinetics
  of receptor-ligand interactions using binding-unbinding correlation
  spectroscopy}}.
\newblock {\emph{\JournalTitle{J. Biol. Chem.}}}
  \textbf{\bibinfo{volume}{287}}, \bibinfo{pages}{35272}
  (\bibinfo{year}{2012}).

\bibitem{LitvinovJBC16}
\bibinfo{author}{Litvinov, R.~I.}, \bibinfo{author}{Farrell, D.~H.},
  \bibinfo{author}{Weisel, J.~W.} \& \bibinfo{author}{Bennett, J.~S.}
\newblock \bibinfo{journal}{\bibinfo{title}{The platelet integrin
  $\alpha${II}b$\beta$3 differentially interacts with fibrin versus
  fibrinogen}}.
\newblock {\emph{\JournalTitle{J. Biol. Chem.}}}
  \textbf{\bibinfo{volume}{291}}, \bibinfo{pages}{7858--7867}
  (\bibinfo{year}{2016}).

\bibitem{MedvedJTH09}
\bibinfo{author}{Medved, L.} \& \bibinfo{author}{Weisel, J.~W.}
\newblock \bibinfo{journal}{\bibinfo{title}{Recommendations for nomenclature on
  fibrinogen and fibrin.}}
\newblock {\emph{\JournalTitle{J. Thromb. Haemost.}}}
  \textbf{\bibinfo{volume}{7}}, \bibinfo{pages}{355--359}
  (\bibinfo{year}{2009}).

\bibitem{EverseBiochemistry98}
\bibinfo{author}{Everse, S.~J.}, \bibinfo{author}{Spraggon, G.},
  \bibinfo{author}{Veerapandian, L.}, \bibinfo{author}{Riley, M.} \&
  \bibinfo{author}{Doolittle, R.~F.}
\newblock \bibinfo{journal}{\bibinfo{title}{Crystal structure of fragment
  double-d from human fibrin with two different bound ligands}}.
\newblock {\emph{\JournalTitle{Biochemistry}}} \textbf{\bibinfo{volume}{37}},
  \bibinfo{pages}{8637--8642} (\bibinfo{year}{1998}).

\bibitem{Charmm09}
\bibinfo{author}{Brooks, B.~R.} \emph{et~al.}
\newblock \bibinfo{journal}{\bibinfo{title}{{CHARMM}: The biomolecular
  simulation program}}.
\newblock {\emph{\JournalTitle{J. Comput. Chem.}}}
  \textbf{\bibinfo{volume}{30}}, \bibinfo{pages}{1545--1614}
  (\bibinfo{year}{2009}).

\bibitem{YeeStructure97}
\bibinfo{author}{Yee, V.~C.} \emph{et~al.}
\newblock \bibinfo{journal}{\bibinfo{title}{Crystal structure of a 30 k{D}a
  {C}-terminal fragment from the $\gamma$-chain of human fibrinogen.}}
\newblock {\emph{\JournalTitle{Structure}}} \textbf{\bibinfo{volume}{15}},
  \bibinfo{pages}{125--138} (\bibinfo{year}{1997}).

\bibitem{KostelanskyBiochem02}
\bibinfo{author}{Kostelansky, M.~S.}, \bibinfo{author}{Betts, L.},
  \bibinfo{author}{Gorkun, O.~V.} \& \bibinfo{author}{Lord, S.~T.}
\newblock \bibinfo{journal}{\bibinfo{title}{{2.8 \AA} crystal structures of
  recombinant fibrinogen fragment {D} with and without two peptide ligands:
  {GHRP} binding to the ``b'' site disrupts its nearby calcium-binding site}}.
\newblock {\emph{\JournalTitle{Biochemistry}}} \textbf{\bibinfo{volume}{41}},
  \bibinfo{pages}{12124--12132} (\bibinfo{year}{2002}).

\bibitem{KononovaJBC13}
\bibinfo{author}{Kononova, O.} \emph{et~al.}
\newblock \bibinfo{journal}{\bibinfo{title}{Molecular mechanisms,
  thermodynamics, and dissociation kinetics of knob-hole interactions in
  fibrin}}.
\newblock {\emph{\JournalTitle{J. Biol. Chem.}}}
  \textbf{\bibinfo{volume}{288}}, \bibinfo{pages}{22681--22692}
  (\bibinfo{year}{2013}).

\bibitem{FerraraProteins02}
\bibinfo{author}{Ferrara, P.}, \bibinfo{author}{Apostolakis, J.} \&
  \bibinfo{author}{Caflisch, A.}
\newblock \bibinfo{journal}{\bibinfo{title}{Evaluation of a fast implicit
  solvent model for molecular dynamics simulations}}.
\newblock {\emph{\JournalTitle{Proteins}}} \textbf{\bibinfo{volume}{46}},
  \bibinfo{pages}{24--33} (\bibinfo{year}{2002}).

\bibitem{ZhmurovJACS12}
\bibinfo{author}{Zhmurov, A.} \emph{et~al.}
\newblock \bibinfo{journal}{\bibinfo{title}{Mechanical transition from
  $\alpha$-helical coiled coils to $\beta$-sheets in fibrin(ogen)}}.
\newblock {\emph{\JournalTitle{J. Am. Chem. Soc.}}}
  \textbf{\bibinfo{volume}{134}}, \bibinfo{pages}{20396--20402}
  (\bibinfo{year}{2012}).

\bibitem{KononovaBiochem17}
\bibinfo{author}{Kononova, O.} \emph{et~al.}
\newblock \bibinfo{journal}{\bibinfo{title}{Mechanistic basis for the binding
  of {RGD}-and {AGDV}-peptides to the platelet integrin
  {$\alpha$IIb$\beta$3}}}.
\newblock {\emph{\JournalTitle{Biochemistry}}} \textbf{\bibinfo{volume}{56}},
  \bibinfo{pages}{1932--1942} (\bibinfo{year}{2017}).

\end{thebibliography}
